\documentclass[
  aps,
  pra,
  reprint,
  nofootinbib,
  floatfix,
  superscriptaddress
]{revtex4-2}
\usepackage[T1]{fontenc}
\usepackage[utf8]{inputenc}
\usepackage{lmodern}
\usepackage{amsmath,amssymb}
\usepackage{bm}
\usepackage{graphicx}
\newsavebox{\panelgraphicbox}
\newcommand{\panelgraphic}[3][]{%
  \begingroup
  \sbox{\panelgraphicbox}{\includegraphics[#1]{#2}}%
  \setlength{\unitlength}{0.01\wd\panelgraphicbox}%
  \usebox{\panelgraphicbox}%
  \makebox[0pt][r]{%
    \begin{picture}(100,0)%
      #3%
    \end{picture}%
  }%
  \endgroup
}
\newcommand{\panellabel}[3]{%
  \put(#1,#2){\makebox(0,0)[lt]{\normalfont\small\bfseries (#3)}}%
}
\usepackage[protrusion=true,expansion=false]{microtype}
\usepackage{placeins}
\graphicspath{{figures/}}
\newcommand{\ket}[1]{\lvert #1\rangle}
\newcommand{\bra}[1]{\langle #1\rvert}
\usepackage{xcolor}

\begin{document}
\title{Weighted hypergraph states as a resource for quantum metrology}
\author{B. J. Alexander}
\email{alexander.byronj@gmail.com}
\affiliation{Department of Physics, Stellenbosch University, Matieland 7602, South Africa}
\author{M. S. Tame}
\affiliation{Department of Physics, Stellenbosch University, Matieland 7602, South Africa}
\date{\today}

\begin{abstract}
Quantum correlations can be used to improve sensing precision beyond classical limits, yet the correlated resource states needed can be challenging to prepare and preserve. Here, we study the family of weighted hypergraph states and identify subclasses characterized by their underlying graph topology that exhibit enhanced metrological performance. Using the quantum Fisher information and a novel Dicke state basis representation, we show that these subclasses outperform both separable and standard weighted graph state benchmarks over broad ranges of edge weights corresponding to weak controlled operations between qubits representing the vertices of the graph. We provide analytical formulas for an arbitrary number \(N\) of qubits, supplemented by a general study, while focusing a detailed analysis on the example case of \(N=4\). By identifying the critical probabilities for local dephasing, we show that adding higher-order weighted hyperedges for $N$ qubits can enhance metrological robustness, allowing the resulting states to retain a quantum advantage at higher noise levels. These probe states exhibit favorable scaling, enhanced precision, and noise tolerance. Our findings establish weighted hypergraph states as viable, structurally tunable and robust resources for quantum-enhanced metrology.
\end{abstract}

\maketitle

\section{Introduction}
Quantum metrology uses controllable quantum systems and nonclassical resources, including multipartite entanglement~\cite{horodecki2009quantum}, to enhance precision in parameter estimation~\cite{giovannetti2011advances,toth2014quantum,liu2020quantum,degen2017quantum,pezze2018quantum,montenegro2025review}. Under comparable encoding and resource constraints, suitable entangled probe states can achieve precision beyond that of separable states~\cite{toth2014quantum}.
Canonical resources for quantum-enhanced metrology include maximally entangled Greenberger--Horne--Zeilinger (GHZ) states~\cite{toth2014quantum,greenberger1989going,bollinger1996optimal}, NOON states~\cite{toth2014quantum,boto2000quantum}, and spin-squeezed states~\cite{toth2014quantum,wineland1992spin,kitagawa1993squeezed}.

While quantum entanglement can enhance measurement precision, many highly entangled probe states, including GHZ states, are particularly sensitive to decoherence in real physical systems~\cite{huelga1997improvement}. Under standard assumptions, generic uncorrelated noise can restore classical asymptotic scaling while still permitting constant-factor improvements~\cite{huelga1997improvement,demkowicz2012elusive}.
The practical objective in experiments is therefore to combine enhanced sensing using finite resources with robustness against realistic noise.
Approaches developed toward this goal include optimized many-body states~\cite{pezze2018quantum}, spin-squeezed states~\cite{wineland1992spin,kitagawa1993squeezed},
projected squeezed-state protocols~\cite{alexander2020generating,alexander2024robustness}, interaction-based readout using time-reversal dynamics~\cite{linnemann2016quantum}, decoherence-free sensing~\cite{dorner2012quantum, roos2006designer}, and quantum-error-corrected metrology~\cite{kessler2014quantum,arrad2014increasing, dur2014improved,ozeri2013heisenberg}.

Experiments across diverse physical platforms have demonstrated quantum-enhanced sensing and some of the above-mentioned advances in state preparation, control, and readout.
Trapped-ion experiments have approached Heisenberg-limited spectroscopy~\cite{leibfried2004toward}, demonstrated enhanced displacement and electric-field sensing~\cite{gilmore2021quantum}, and realized programmable near-optimal sensors~\cite{marciniak2022optimal}.
Neutral-atom platforms have demonstrated interferometry beyond classical precision limits~\cite{gross2010nonlinear} and the preparation and readout of entangled probe states in optical clocks~\cite{finkelstein2024universal}.
Photonic experiments include four-photon phase measurements~\cite{nagata2007beating} and phase estimation beyond the shot-noise limit~\cite{slussarenko2017unconditional}.
Superconducting qubits have enabled metrological characterization of non-Gaussian states~\cite{xu2022metrological}. Furthermore, nitrogen-vacancy ensembles have shown improved sensing through robust control~\cite{zhou2020quantum} and signal amplification through time-reversed many-body dynamics~\cite{gao2025signal}.

 In this work, we study weighted hypergraph states as resources for quantum metrology. Here, there is the possibility of using tunable weights to gain flexibility in state preparation while higher-order quantum correlations from hypergraph structures may enhance robustness to noise. Previous studies have shown that certain fully weighted hypergraph states can achieve Heisenberg-limited precision~\cite{gachechiladze2016extreme}. Our study explores the more general class of weighted hypergraph states. In doing so, we extend weighted graph state metrology~\cite{alexander2026weighted,skosana2026experimental} to weighted hypergraph states. Building on the established utility and robustness of $N$-qubit weighted complete and star graph states~\cite{shettell2020graph,alexander2026weighted}, we consider weighted graph states with higher-order hyperedge layers that are symmetric under permutations of the leaf qubits. Classifying these probe states by their hyperedge-orbit structure, we evaluate the quantum Fisher information (QFI) against separable and standard weighted graph state benchmarks, finding that selected subclasses outperform both over substantial interaction-weight intervals below the fully weighted value of \(\phi=\pi\). Furthermore, under dephasing, critical-probability evaluations confirm that hyperedge structures maintain a quantum advantage with improved robustness. 

 This work adds to the growing interest in using hypergraph states as resources in quantum information processing. Entanglement classification, entanglement witnesses, and the geometric measure of entanglement for hypergraph states are discussed in Ref.~\cite{gachechiladze2019thesis}. A broader overview of their structure, stabilizer formalism, and applications is provided in Ref.~\cite{salem2025review}. Recent studies connect hypergraph structure to bipartite entanglement through Schmidt-rank certificates~\cite{fajardo2026restriction} and characterize multipartite entanglement through statistical-mechanical mappings~\cite{scarafile2026statistical}. Related edge-ordered hypergraph constructions connect hypergraph properties to entanglement~\cite{susulovska2026three}. The entanglement of randomized hypergraph states under probabilistic hyperedge generation has been studied~\cite{salem2026randomized}, while protocols for noise characterization have been developed for graph and hypergraph states~\cite{park2026lowoverhead}.

The paper is structured as follows: Section~II provides definitions and describes the probe state families and phase encoding; Section~III presents the analytic QFI expressions and dephasing results; Section~IV discusses the results. Appendices~A--D provide the derivations and numerical details.
\begin{figure*}[!t]
    \centering
    \includegraphics[width=0.92\textwidth]{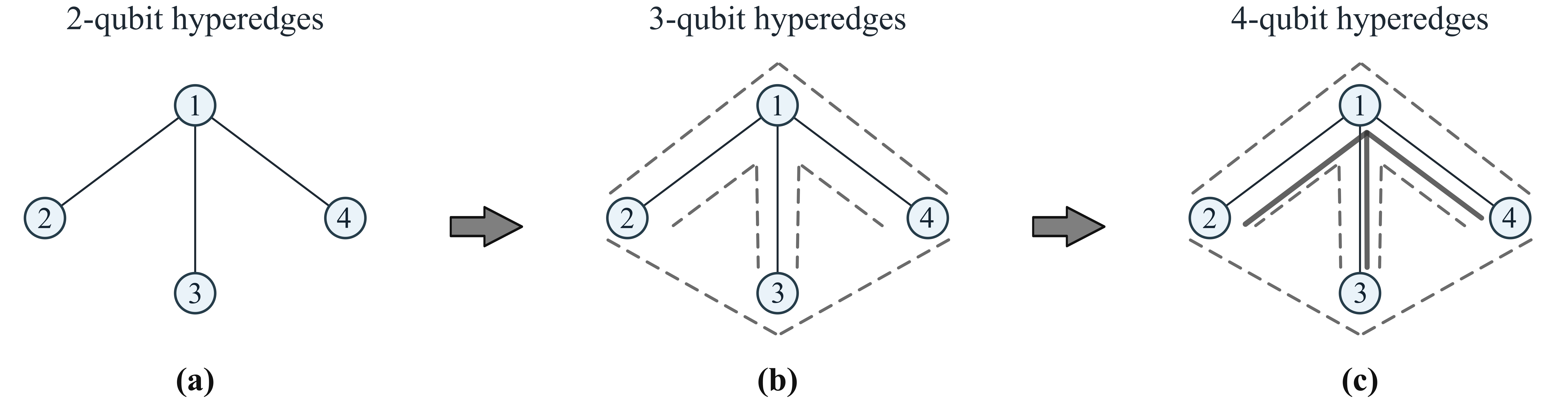}
    \caption{Preparation of a four-qubit weighted star hypergraph state containing all higher-order hyperedges. (a) The weighted star graph formed by the pairwise edges \(\{1,2\},\{1,3\},\{1,4\}\), shown as thin solid segments. (b) Addition of all four three-qubit hyperedges, shown as dashed contours. (c) Subsequent addition of the four-qubit hyperedge \(\{1,2,3,4\}\), shown as the thicker solid branched contour. All hyperedges carry the same phase weight \(\phi_e=\phi\); bends and crossings are purely graphical and do not represent additional qubits. The controlled unitary operation associated with each hyperedge is given in the main text.}
    \label{fig:hypergraph-progression}
\end{figure*}
\section{Theory}

A metrological protocol infers an unknown parameter from measurement outcomes. For a chosen measurement, the classical Fisher information (CFI) quantifies how sensitively the outcome probabilities depend on that parameter. On the other hand, the QFI is the maximum CFI over all possible measurements and characterizes the best sensitivity available from the encoded probe. Therefore, the QFI sets the precision limit for a given probe state and parameter encoding; attaining this limit requires a measurement whose CFI equals the QFI~\cite{jacobs2014quantum,toth2014quantum,liu2020quantum}.

More formally, a probe state \(\rho\) acquires an unknown phase \(\theta\) through the unitary encoding
\begin{equation}
   \rho_{\theta}
    =
    \hat{U}_{\theta}\rho\hat{U}_{\theta}^{\dagger},
    \qquad
    \hat{U}_{\theta}
    =
    \exp(-i\theta\hat{H}_{\mathrm{enc}}),
    \label{Uni}
\end{equation}
where \(\hat{H}_{\mathrm{enc}}\) is the Hermitian encoding generator. For a pure state, the QFI is
\begin{equation}
    \mathcal{F}_{Q}
    =
    4\left(
        \langle\hat{H}_{\mathrm{enc}}^{2}\rangle
        -
        \langle\hat{H}_{\mathrm{enc}}\rangle^{2}
    \right)
    =
    4(\Delta H_{\mathrm{enc}})^{2},
    \label{FQH}
\end{equation}
where \(\Delta H_{\mathrm{enc}}\) denotes the standard deviation of the generator for the probe state. The quantum Cram\'er--Rao bound
\begin{equation}
    (\Delta \theta_{\mathrm{est}})^2 \geq \frac{1}{\nu\mathcal{F}_{Q}},
\end{equation}
relates the QFI to the variance \((\Delta\theta_{\mathrm{est}})^2\) of a locally unbiased estimator \(\hat{\theta}_{\mathrm{est}}\), with \(\nu\) independent measurements~\cite{giovannetti2011advances,toth2014quantum,liu2020quantum}.
For an \(N\)-qubit local linear generator whose single-qubit terms have eigenvalues \(\pm1/2\), separable probes satisfy \(\mathcal{F}_{Q}\leq N\), defining the standard quantum limit (SQL). Entangled probes can exceed this bound and reach \(\mathcal{F}_{Q}=N^2\), the Heisenberg limit (HL), under ideal noiseless conditions~\cite{giovannetti2011advances,toth2014quantum}. These bounds provide benchmarks for the probe state families below.

Graph states can be used as probe states for quantum metrology~\cite{shettell2020graph}; they associate qubits with vertices and pairwise controlled-\(\hat{Z}\) (CZ) gates with edges~\cite{hein2006entanglement}. They are prepared from an initial product state \(\ket{+}^{\otimes N}\), where \(\ket{+}=(\ket{0}+\ket{1})/\sqrt{2}\).
Changing the connectivity via edges can change the correlations between qubits, making the graph a convenient way to organize and compare candidate probes. Their metrological performance depends on graph structure and many-body correlations~\cite{friis2017flexible,shettell2020graph,nguyen2024harnessing,plodzien2025manybody}. Related graph state resources have been explored in atomic
ensembles~\cite{cooper2024graph}, Rydberg
arrays~\cite{qin2025scaling}, and photonic
systems~\cite{li2026deterministic}.

Weighted graph states are more general and replace the fixed CZ gates with controlled-phase gates with tunable weight \(\phi\), recovering standard graph states at \(\phi=\pi\)~\cite{alexander2026weighted}. Such states arise from Ising-type interactions~\cite{hartmann2007weighted}. The weight $\phi$ specifies the strength of each controlled-phase operation; it is a preparation parameter, distinct from the unknown phase \(\theta\) to be estimated. Theory has shown metrological advantages at reduced weights~\cite{alexander2026weighted}, and a two-qubit photonic proof-of-concept has been demonstrated~\cite{skosana2026experimental}. Fusion-based preparation has also been proposed~\cite{rimock2026type}.

Hypergraph states extend the above construction to gates acting jointly on more than two qubits~\cite{rossi2013quantum,guhne2014entanglement,poderini2026hypergraph}. Recent work explicitly expresses the stabilizers of complete hypergraph states as linear combinations of tensor products of Pauli operators~\cite{arantes2026stabilizers}. Weighted hypergraph states assign a tunable phase \(\phi_e\) to each hyperedge~\cite{zakaryan2025nonsymmetric}, providing additional control over interaction order, connectivity, and strength. The hypergraph describes which subsets participate in these operations, while the edge weights describe how strongly their joint phases are modified. Four-qubit fully weighted hypergraph states have been prepared on a silicon-photonic chip~\cite{huang2024demonstration}, and three-qubit fully weighted examples have been studied for metrology~\cite{bag2022achieving}. Here, we go beyond these works and study weighted hypergraph states as probe states for quantum metrology. 

Fig.~\ref{fig:hypergraph-progression} illustrates the construction of an example weighted hypergraph state with four qubits: pairwise edges form the weighted star in panel~(a), three-qubit hyperedges are added in panel~(b), and the four-qubit hyperedge is added in panel~(c). The precise weighted hypergraph state definitions and probe families are introduced below.

\subsection{Weighted hypergraph state probes}

Let \(\mathcal{H}=(V,E)\) be a hypergraph with vertex set \(V=\{1,\ldots,N\}\) and hyperedge set \(E\), where each \(e\in E\) satisfies \(e\subseteq V\).
For each \(e\in E\), define the projector that selects the
all-ones configuration on \(e\) and acts as the identity elsewhere:
\begin{equation}
\hat{\Pi}_{e}=\bigotimes_{j=1}^{N}\hat{P}_{j}^{(e)},
\qquad
\hat{P}_{j}^{(e)}=
\begin{cases}
\ket{1}_{j}\!\bra{1}, & j\in e,\\
\hat{\mathbb{I}}_{j}, & j\notin e
\end{cases},
\end{equation}
where \(\hat{\mathbb{I}}_{j}\) denotes the identity operator on vertex
\(j\). The corresponding weighted controlled-phase operation is defined as
\begin{equation}
\hat{C}_{e}(\phi_{e})
=
\exp\!\left(i\phi_{e}\hat{\Pi}_{e}\right)
=
\hat{\mathbb{I}}
+
\left(e^{i\phi_{e}}-1\right)\hat{\Pi}_{e},
\label{eq:weighted-hyperedge-gate}
\end{equation}
where \(\hat{\mathbb{I}}\) denotes the identity on the
full \(N\)-qubit Hilbert space.
The standard fully weighted hyperedge phase gate is recovered at
\(\phi_{e}=\pi\). Since all operations in
Eq.~\eqref{eq:weighted-hyperedge-gate} are diagonal in the computational
basis, they commute, and the weighted hypergraph state is defined as
\begin{equation}
    \ket{\mathcal{H}(\{\phi_e\})}
    =
    \prod_{e\in E}\hat{C}_{e}(\phi_{e})
    \ket{+}^{\otimes N}.
    \label{eq:weighted-hypergraph-state}
\end{equation}

Throughout this study, we consider uniformly weighted hypergraph states, with \(\phi_e=\phi\in[0,2\pi)\) for every \(e\in E\).  This restriction
provides a common interaction-strength parameter for comparing hyperedge structures while retaining a continuously
tunable interaction strength, in direct analogy with the uniformly weighted
graphs studied in Ref.~\cite{alexander2026weighted}. Among the connected weighted graph topologies examined, the star and complete families demonstrate superior parameter estimation
precision over particular weight intervals. Consequently, here we focus primarily on star-based weighted hypergraph states, whose pair-edge backbone requires only $N-1$ controlled-phase interactions, compared with $N(N-1)/2$ for the complete graph, and thus scales linearly rather than quadratically with $N$. We also consider the complete weighted hypergraph as a fully symmetric benchmark. 

For the weighted star hypergraph states, let qubit \(1\) be the central qubit and
$L=\{2,\ldots,N\}$ denote the set of leaf qubits. The star-based orbit-layer families studied below share the weighted star graph’s pair-edge backbone,
\begin{equation}
    E_{\star}
    =
    \bigl\{\{1,j\}:j\in L\bigr\}.
    \label{eq:star-backbone}
\end{equation}
For the symmetric subclasses considered here, we add higher-order hyperedges in complete orbits under leaf permutations. An orbit is the set of all distinct hyperedges obtained from a given hyperedge by permuting the leaves while keeping the central vertex fixed. The central and leaf-only orbits of \(r\)-body hyperedges are,
respectively,
\begin{align}
    A_r
    &=\bigl\{e\subseteq V:|e|=r,\ 1\in e\bigr\},
    \label{eq:general-orbits}
    \\
    B_r
    &=\bigl\{e\subseteq L:|e|=r\bigr\}.
    \label{eq:general-orbits2}
\end{align}
Thus, \(A_r\) contains all \(r\)-qubit hyperedges that include the
central vertex, while \(B_r\) contains all \(r\)-qubit hyperedges
formed entirely from leaves.
The allowed orders are \(r=3,\ldots,N\) for \(A_r\) and
\(r=3,\ldots,N-1\) for \(B_r\).
With uniform edge weights, including or excluding complete orbits
ensures that each resulting probe is invariant under leaf permutations.

\begin{table*}[t]
    \caption{The eight orbit signatures for the four-qubit weighted
    star hypergraph state.  The entries \((a_3,b_3,a_4)\) specify inclusion
    of the complete orbits \(A_3,B_3,A_4\), respectively, giving
    \(n_{\mathrm{add}}=3a_3+b_3+a_4\) added higher-order hyperedges.
    The final column lists these added hyperedges;
    the star backbone \(E_{\star}\) is common to all states, and all active
    edges carry the uniform weight \(\phi\).}
    \label{tab:orbit-layer-signatures}
    \centering
    \squeezetable
    \begin{ruledtabular}
    \begin{tabular}{ccl}
        \shortstack{Orbit signature\\
        \(\bm{\sigma}_4=(a_3,b_3,a_4)\)}
        & \shortstack{Number of added\\higher-order hyperedges}
        & Added higher-order hyperedges \\
        \colrule
        \((0,0,0)\) & \(0\) & \textbf{none} \\
        \((0,0,1)\) & \(1\) & \(\{1,2,3,4\}\) \\
        \((0,1,0)\) & \(1\) & \(\{2,3,4\}\) \\
        \((0,1,1)\) & \(2\) & \(\{2,3,4\},\ \{1,2,3,4\}\) \\
        \((1,0,0)\) & \(3\) & \(\{1,2,3\},\ \{1,2,4\},\
        \{1,3,4\}\) \\
        \((1,0,1)\) & \(4\) & \(\{1,2,3\},\ \{1,2,4\},\
        \{1,3,4\},\ \{1,2,3,4\}\) \\
        \((1,1,0)\) & \(4\) & \(\{1,2,3\},\ \{1,2,4\},\
        \{1,3,4\},\ \{2,3,4\}\) \\
        \((1,1,1)\) & \(5\) & \(\{1,2,3\},\ \{1,2,4\},\
        \{1,3,4\},\ \{2,3,4\},\ \{1,2,3,4\}\) \\
    \end{tabular}
    \end{ruledtabular}
\end{table*}

Fig.~\ref{fig:hypergraph-progression} illustrates an example of this construction for
\(N=4\), with the weighted star graph successively supplemented by all
three-qubit hyperedges and then the four-qubit hyperedge. The available
higher-order hyperedge orbits are
\begin{align}
    A_3
    &=
    \bigl\{\{1,2,3\},\{1,2,4\},\{1,3,4\}\bigr\},
    \\
    B_3
    &=
    \bigl\{\{2,3,4\}\bigr\},
    \\
    A_4
    &=
    \bigl\{\{1,2,3,4\}\bigr\}.
\end{align}
We label the eight leaf-permutation-symmetric probes by the binary orbit signature
\begin{equation}
    \bm{\sigma}_4=(a_3,b_3,a_4)\in\{0,1\}^{3},
    \label{eq:sigma4}
\end{equation}
where each entry specifies whether the corresponding complete orbit
\(A_3\), \(B_3\), or \(A_4\) is included. We denote the union of the
selected orbits by \(\mathcal{E}^{(>2)}_{\bm{\sigma}_4}\). For the four-qubit star-based families, the full edge set used in Eq.~\eqref{eq:weighted-hypergraph-state} is \(E=E_{\star}\cup\mathcal{E}^{(>2)}_{\boldsymbol{\sigma}_4}\), with the eight orbit signatures and their corresponding added hyperedges listed in Table~\ref{tab:orbit-layer-signatures}.

We use \(N=4\) as our primary exploratory example because it is the smallest star-based system that supports both center-containing and leaf-only three-qubit hyperedges, as well as a four-qubit hyperedge. The three independently selectable higher-order orbits provide a richer range of hyperedge configurations and associated multipartite entanglement structures for metrological comparison, while retaining the four-qubit setting used for the weighted star and complete graph examples in Ref.~\cite{alexander2026weighted}.

The orbit construction extends to arbitrary \(N\) within the same family of probes
invariant under leaf permutations. For the four-qubit system, write a computational-basis state as
\(\ket{x}=\ket{c}\ket{y}\), where \(x=(c,x_2,x_3,x_4)\)
is the full bit string, \(c=x_1\in\{0,1\}\) is the central bit, and
\(y=(x_2,x_3,x_4)\in\{0,1\}^3\) specifies the three leaves.
The Hamming weight
\(w=|y|=x_2+x_3+x_4\in\{0,1,2,3\}\)
specifies how many of the three leaf qubits are in \(\ket{1}\),
while \(|x|=c+w\).
Any two leaf bit strings with the same Hamming weight contain the same numbers of zeros and ones, so a permutation of the leaves maps one to the other. Since each included orbit in Eqs.~\eqref{eq:general-orbits}
and~\eqref{eq:general-orbits2} is complete and uniformly weighted,
leaf permutations map its hyperedges to other included hyperedges
with the same weights. This merely reorders the commuting phase
gates in Eq.~\eqref{eq:weighted-hyperedge-gate}, so the full
preparation unitary commutes with these permutations. Together with the symmetry of the initial state
\(\ket{+}^{\otimes4}\), this implies that the amplitudes
of the prepared state given by
Eq.~\eqref{eq:weighted-hypergraph-state} depend only
on the central bit \(c\) and the leaf Hamming weight \(w\).  

The \(\binom{3}{w}\) leaf bit strings of Hamming weight \(w\) can therefore
be grouped into the normalized Dicke state~\cite{dicke1954coherence}
\begin{equation}
    \ket{D_w^3}
    =
    \binom{3}{w}^{-1/2}
    \sum_{|y|=w}\ket{y},
    \label{eq:dicke-state}
\end{equation}
where the sum includes every leaf bit string with exactly \(w\)
ones. The normalization follows from the orthogonality of the
\(\binom{3}{w}\) computational-basis states in the sum.
The four states \(\ket{D_w^3}\), with \(w=0,1,2,3\), form a basis
for the permutation-symmetric subspace of the leaves. To determine the phases in the Dicke-basis expansion, consider each computational-basis state $\ket{c}\ket{y}$ separately. In $\ket{+}^{\otimes 4}$, the coefficient of every such state is $1/4$; the diagonal preparation gates change its phase but not its magnitude.
Specifically, each included hyperedge gate
multiplies a component's amplitude by \(e^{i\phi}\) when every
qubit in that hyperedge is in \(\ket{1}\), and by unity otherwise.
Since all included gates have the same weight \(\phi\), the total
phase is determined by the number of included hyperedges whose
vertices all carry the bit value one.

The star pair edges contribute \(cw\): each of the \(w\) leaves
in \(\ket{1}\) contributes one phase factor when \(c=1\).
For the central three-qubit orbit \(A_3\), any two of these leaves
can combine with the central qubit, giving
\(a_3c\binom{w}{2}\) contributions.
The leaf-only orbit \(B_3\) requires three leaves in \(\ket{1}\),
giving \(b_3\binom{w}{3}\).
The four-qubit orbit \(A_4\) additionally requires \(c=1\),
giving \(a_4c\binom{w}{3}\).
Thus, the total number of $e^{i \phi}$ phase contributions is
\begin{equation}
    g_{\bm{\sigma}_4}(c,w)
    =
    cw
    +a_3c\binom{w}{2}
    +b_3\binom{w}{3}
    +a_4c\binom{w}{3},
    \label{eq:n4-phase-count}
\end{equation}
where \(a_3,b_3,a_4\in\{0,1\}\) indicate whether the respective
orbits are included. Binomial coefficients with a lower index
greater than \(w\) are understood to vanish. Every computational-basis component with the same \(c\) and \(w\)
therefore acquires the common phase factor
\(e^{i\phi g_{\bm{\sigma}_4}(c,w)}\).
Grouping the \(\binom{3}{w}\) equal-amplitude components into a
normalized Dicke state gives the coefficient
\(\sqrt{\binom{3}{w}}/4\).
The resulting probe is
\begin{equation}
    \ket{\psi_{\bm{\sigma}_4}(\phi)}
    =
    \frac{1}{4}
    \sum_{c=0}^{1}\sum_{w=0}^{3}
    \sqrt{\binom{3}{w}}\,
    e^{i\phi g_{\bm{\sigma}_4}(c,w)}
    \ket{c}\ket{D_w^3}.
    \label{eq:n4-dicke-state}
\end{equation}

For example, consider \(\bm{\sigma}_4=(1,1,1)\), for which all
three higher-order orbits are included, and the computational-basis
component \(\ket{1101}\).
Here \(c=1\) and \(y=(1,0,1)\), giving \(w=2\).
The pair edges \(\{1,2\}\) and \(\{1,4\}\), together with the
central three-qubit hyperedge \(\{1,2,4\}\), each contribute a
factor \(e^{i\phi}\). All other included hyperedges contain
qubit \(3\), whose bit value is zero.
Accordingly, Eq.~\eqref{eq:n4-phase-count} gives $g_{(1,1,1)}(1,2)=3$.
The component therefore acquires the phase factor \(e^{3i\phi}\).
The other two components with \(c=1\) and \(w=2\), namely
\(\ket{1110}\) and \(\ket{1011}\), acquire the same factor:
although the contributing hyperedges have different labels,
their counts remain two star pairs and one central triple.
Since each component initially has amplitude \(1/4\), their
contributions combine as
\begin{equation}
    \begin{aligned}
        &\frac{e^{3i\phi}}{4}
        \left(\ket{1110}+\ket{1101}+\ket{1011}\right)\\
        &\qquad=
        \frac{\sqrt{3}}{4}e^{3i\phi}
        \ket{1}\ket{D_2^3}.
    \end{aligned}
\end{equation}
This is precisely the \(c=1,w=2\) term in the Dicke-basis
expansion of Eq.~\eqref{eq:n4-dicke-state}.

The ideal star-based probes considered here are invariant under
permutations of the leaf qubits.
For \(N=4\), tensor products of the two central-qubit basis states
($|0\rangle$ and $|1\rangle$) with the four leaf Dicke states span an
eight-dimensional subspace of the full 16-dimensional Hilbert space.
For general \(N\), the symmetric subspace of the \(N-1\) leaves has
\(N\) Dicke basis states (\(w=0,\ldots,N-1\)), giving a
\(2N\)-dimensional representation in the product basis
\(\lvert c\rangle\lvert D_w^{N-1}\rangle\), with \(c\in\{0,1\}\)
(see Appendix~\ref{app:symmetric-representation}).

In contrast, an \(N\)-qubit uniformly weighted complete hypergraph
state is invariant under permutations of all \(N\) qubits. It lies
in the fully symmetric subspace spanned by the Dicke states
\(\lvert D_s^N\rangle\), with \(s=0,\ldots,N\), giving an
\((N+1)\)-dimensional representation.

These reductions apply to the ideal pure probes. Independent and
identically distributed (i.i.d.) phase-flip noise preserves
the corresponding permutation invariance of the density operator
but can populate sectors outside these subspaces, so noisy probes
need not remain supported on these reduced subspaces.

\subsection{Encoding and reference probes}

For the star-based probes, the unknown phase \(\theta\) is encoded with a
generator that preserves leaf-permutation symmetry,
\begin{equation}
    \hat{H}_{\star}\equiv\hat{H}_{\mathrm{enc}}
    =
    \frac{1}{2}
    \left(
        \hat{Z}_{1}
        +
        \sum_{j=2}^{N}\hat{X}_{j}
    \right).
    \label{eq:encoding-generator}
\end{equation}
Here, \(\hat{X}_{j}\) and \(\hat{Z}_{j}\) are the Pauli operators acting on qubit \(j\).
For the complete-hypergraph comparison, we use the homogeneous generator
\begin{equation}
    \hat{H}_{\mathrm{comp}}
    =\frac{1}{2}\sum_{j=1}^{N}\hat{X}_{j}.
    \label{eq:complete-encoding-generator}
\end{equation}
The star generator treats the central subsystem (on which $\hat{Z}_{1}$ acts) and the leaf subsystems according to the star
topology and commutes with all permutations of the leaves. With the local normalization used in Eqs.~\eqref{eq:encoding-generator} and \eqref{eq:complete-encoding-generator}, the maximum QFI defined in Eq.~\eqref{FQH} is $N$ for separable states and $N^2$ over all $N$-qubit states. 

Throughout, \(\mathcal{F}_Q\) denotes the QFI generically, while a
probe-family subscript identifies the chosen probe for the stated
encoding generator. For a general $N$-qubit probe
family \(\mathcal{G}\), we write \(\mathcal{F}_{\mathcal{G}}(N,\phi,p)\) for its QFI
after dephasing with probability~\(p\).  At \(N=4\), where a family is
identified by its orbit signature, we abbreviate this as
\(\mathcal{F}_{\bm{\sigma}_4}(\phi,p)\). Because the weighted star contains only two-qubit edges, we
simply denote its orbit signature \((0,0,0)\) by \(\star\). The QFI for
arbitrary \(N\) is given as~\cite{alexander2026weighted}
\begin{equation}
    \begin{aligned}
        \mathcal{F}_{\star}(N,\phi,0)
        ={}&
        \frac{3N^2+4N+1}{8}
        -\frac{N^2-1}{2}\cos\phi
        \\
        &+
        \frac{(N-1)(N-3)}{8}\cos(2\phi).
    \end{aligned}
    \label{eq:star-qfi}
\end{equation}
For comparison, we also consider the complete weighted hypergraph state, which contains every hyperedge
of order two or higher,
\begin{equation}
    E_{\mathrm{comp}}^{(N)}
    =\bigl\{e\subseteq V:2\leq |e|\leq N\bigr\}.
    \label{eq:complete-hypergraph-edges}
\end{equation}
Under \(\hat{H}_{\mathrm{comp}}\), its pure-state QFI is
\begin{equation}
    \begin{aligned}
    \mathcal{F}_{\mathrm{comp}}(N,\phi,0)
    ={}&N+N(N-1)\mathsf{C}_{N}^{\mathrm{comp}}\\
       &-N^2\bigl(\mathsf{A}_{N}^{\mathrm{comp}}\bigr)^2,
    \end{aligned}
    \label{eq:complete-hypergraph-qfi}
\end{equation}
where the dependence of the correlators on \(\phi\) is suppressed and
\begin{align}
    \mathsf{A}_{N}^{\mathrm{comp}}
    &=2^{1-N}\sum_{r=0}^{N-1}\binom{N-1}{r}
      \cos\!\left[\phi(2^r-1)\right],
    \label{eq:complete-hypergraph-A}\\
    \mathsf{C}_{N}^{\mathrm{comp}}
    &=2^{1-N}\sum_{r=0}^{N-2}\binom{N-2}{r}
      \nonumber\\
    &\quad\times
      \left\{1+\cos\!\left[\phi(3\cdot2^r-2)\right]\right\}.
    \label{eq:complete-hypergraph-C}
\end{align}
Appendix~\ref{app:complete-qfi-proof} provides the derivation.  For \(N=4\),
Eq.~\eqref{eq:complete-hypergraph-qfi} yields
\(\mathcal{F}_{\mathrm{comp}}=0, 45/4\) and $4$ at
\(\phi=0, \pi/2\) and $\pi$, respectively.

Later, as an additional benchmark when including dephasing, we also consider the separable state
\begin{equation}
    \ket{\psi_{\mathrm{prod}}}
    =
    \ket{+}_{1}\ket{0}_{2}\cdots\ket{0}_{N},
    \label{eq:product-reference}
\end{equation}
which saturates the SQL in the ideal noiseless case for the generator in
Eq.~\eqref{eq:encoding-generator}. As shown in Appendix~\ref{app:additional-utility}, this state maximizes the QFI at each dephasing probability among all fully separable inputs subjected to local dephasing before encoding with the star generator~$\hat{H}_\star$.

\raggedbottom
\section{Results}

\subsection{Analytic QFI for the hypergraph cutoff family}
\label{sec:analytic-qfi}

\begin{figure*}[!t]
    \centering
    \includegraphics[width=0.63\textwidth]{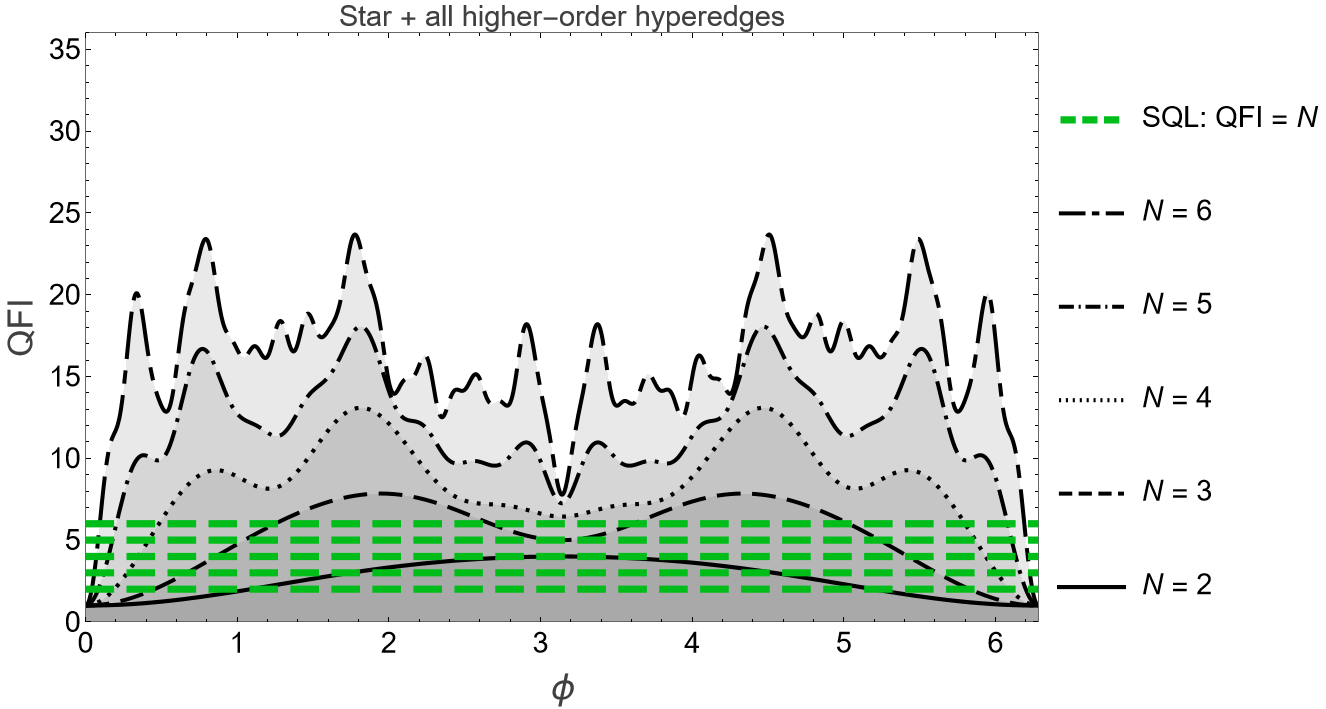}
    \caption{The QFI of the
    weighted star graph state supplemented by all hyperedges of orders \(3,\ldots,N\),
    with the encoding Hamiltonian of Eq.~\eqref{eq:encoding-generator}.
    The black curves, identified by line style in the legend, correspond
    to \(N=2,3,4,5,6\); at \(N=2\), no higher-order hyperedges exist.
    The green dashed horizontal lines show the SQL of
    \(\mathcal{F}_{Q}=N\), one for each \(N\).}
    \label{fig:star-all-qfi-n}
\end{figure*}

\begin{figure*}[!t]
    \centering
    \includegraphics[width=0.63\textwidth]{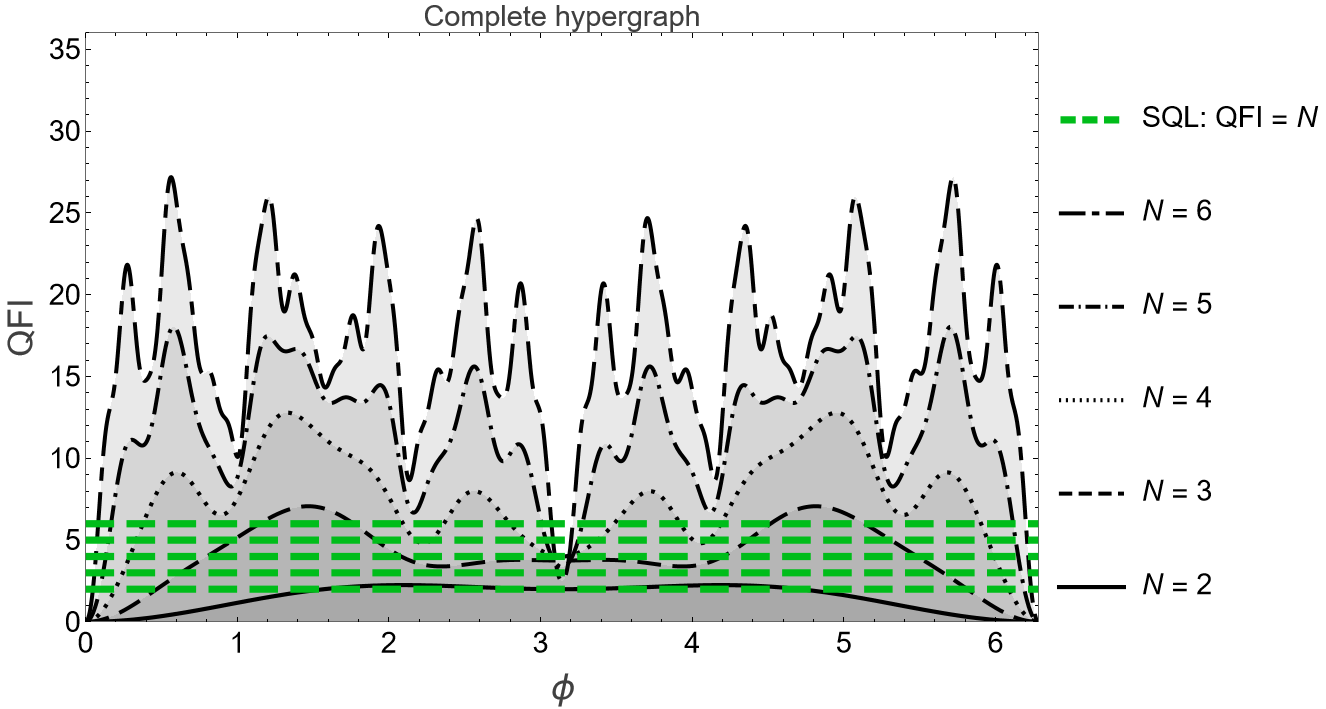}
    \caption{The QFI of
    the complete weighted hypergraph state containing all hyperedges of orders
    \(2,\ldots,N\), with the encoding Hamiltonian of
    Eq.~\eqref{eq:complete-encoding-generator}. The black curves,
    identified by line style in the legend, correspond to
    \(N=2,3,4,5,6\). The green dashed horizontal lines show the SQL of
    \(\mathcal{F}_{Q}=N\), one for each \(N\).}
    \label{fig:complete-qfi-n}
\end{figure*}

\begin{figure*}[!t]
    \centering
    \includegraphics[width=0.9\textwidth]{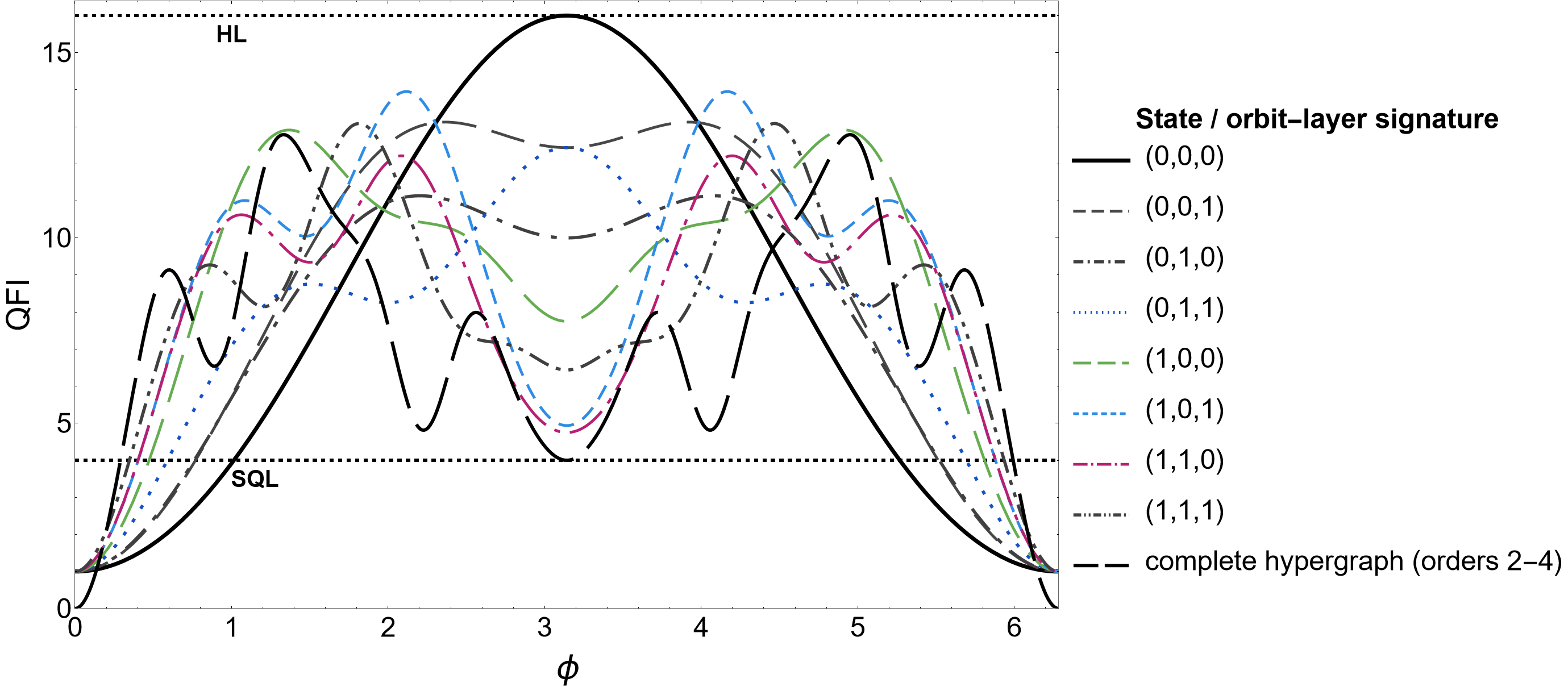}
    \caption{Noiseless QFI for the eight four-qubit weighted star hypergraph
    probes and the complete weighted hypergraph. The binary
    orbit signature \(\bm{\sigma}_4=(a_3,b_3,a_4)\) specifies whether the
    complete orbits \(A_3\), \(B_3\), and \(A_4\) are present on the star backbone. The thick solid black curve represents the weighted star benchmark \(\star=(0,0,0)\), and the long-dashed black curve represents the complete hypergraph containing all hyperedges of orders two through four. The star-based curves use \(\hat{H}_{\star}=(\hat{Z}_1+\hat{X}_2+\hat{X}_3+\hat{X}_4)/2\), whereas the complete-hypergraph curve uses \(\hat{H}_{\mathrm{comp}}=(\hat{X}_1+\hat{X}_2+\hat{X}_3+\hat{X}_4)/2\). The horizontal dotted lines mark the SQL (\(\mathcal{F}_Q=4\)) and HL (\(\mathcal{F}_Q=16\)).}
    \label{fig:symmetric-landscape}
\end{figure*}

We begin with an exact finite-sum expression for the pure-state QFI
of a cutoff family of weighted hypergraph states, formed by adding
all hyperedges of orders \(3,\ldots,k\) to the pair-edge star backbone
in Eq.~\eqref{eq:star-backbone}. For \(N\geq3\) and an integer \(2\leq k\leq N\),
the edge set is
\begin{equation}
    E_{N,k}
    =
    E_{\star}\cup
    \bigl\{e\subseteq V:3\leq |e|\leq k\bigr\}.
    \label{eq:cutoff-edge-set}
\end{equation}
All included edges have the common weight \(\phi\), and
\(\ket{\psi_{N,k}(\phi)}\) is the state in
Eq.~\eqref{eq:weighted-hypergraph-state} with \(E=E_{N,k}\).  The cutoff
\(k=2\) gives the weighted star graph, while \(k=N\) includes every higher-order
hyperedge (a generalization of $\boldsymbol{\sigma}_4 = (1,1,1)$ to $N$ qubits). We write \(\mathcal{F}_{Q}(\phi,k,N)\) for the QFI of the cutoff-family
probe \(\lvert\psi_{N,k}(\phi)\rangle\) under encoding generated by
\(\hat{H}_{\star}\) as defined in Eq.~\eqref{eq:encoding-generator}.

Define the real scalar correlators
\(\mathsf{A}_{N,k}=\langle\hat{X}_2\rangle\),
\(\mathsf{B}_{N,k}=\langle\hat{Z}_1\hat{X}_2\rangle\), and
\(\mathsf{C}_{N,k}=\langle\hat{X}_2\hat{X}_3\rangle\), with expectations
taken in \(\ket{\psi_{N,k}(\phi)}\). Due to the symmetry of the star graph, any leaf or distinct pair of leaves may be used in place of leaves 2 and 3.  Their dependence on \(\phi\) is
suppressed in the notation.  Evaluating the pure-state QFI as \(4\,\operatorname{Var}(\hat H_\star)\)
using leaf symmetry yields
\begin{equation}
    \begin{aligned}
        \mathcal{F}_{Q}(\phi,k,N)
        ={}&N+2(N-1)\mathsf{B}_{N,k}\\
        &+(N-1)(N-2)\mathsf{C}_{N,k}\\
        &-(N-1)^2\mathsf{A}_{N,k}^{\,2},
    \end{aligned}
    \label{eq:cutoff-qfi}
\end{equation}
where 
\begin{align}
    \mathsf{A}_{N,k}
    &=2^{1-N}\sum_{c=0}^{1}\sum_{r=0}^{N-2}
    \binom{N-2}{r}
    \cos\!\left[\phi\Delta_k^{(1)}(c,r)\right],
    \label{eq:cutoff-A}\\
    \mathsf{B}_{N,k}
    &=2^{1-N}\sum_{c=0}^{1}\sum_{r=0}^{N-2}
    (-1)^c\binom{N-2}{r}
    \cos\!\left[\phi\Delta_k^{(1)}(c,r)\right],
    \label{eq:cutoff-B}\\
    \mathsf{C}_{N,k}
    &=2^{1-N}\sum_{c=0}^{1}\sum_{r=0}^{N-3}
    \binom{N-3}{r}
    \left\{1+\cos\!\left[\phi\Delta_k^{(2)}(c,r)\right]\right\}.
    \label{eq:cutoff-C}
\end{align} 
Here, \(c\in\{0,1\}\) is the central bit, and \(r\) denotes the
Hamming weight of the \(N-2\) or \(N-3\) unselected leaf bits
for a single-leaf or two-leaf flip, respectively.
The phase increments are
\begin{align}
    \Delta_k^{(1)}(c,r)
    &=c+D_k(c+r),
    \label{eq:cutoff-delta1}\\
    \Delta_k^{(2)}(c,r)
    &=\Delta_k^{(1)}(c,r)+\Delta_k^{(1)}(c,r+1),
    \label{eq:cutoff-delta2}
\end{align}
with
\begin{equation}
    D_k(s)=\sum_{q=2}^{k-1}\binom{s}{q}.
    \label{eq:cutoff-D}
\end{equation}
The derivations are given in Appendix~\ref{app:analytic-qfi}.
For \(\phi=0\), every cutoff \(k\) produces
\(\ket{+}^{\otimes N}\), so \(\mathcal F_Q(0,k,N)=1\).
Each leaf qubit is in an eigenstate of \(\hat X_j\) and contributes zero
variance; only the central \(\hat Z_1\) term contributes. At \(k=2\), Eq.~\eqref{eq:star-qfi} and the optimal star-graph limit \(\mathcal F_Q(\pi,2,N)=N^2\) under \(\hat H_\star\) are recovered~\cite{alexander2026weighted}.

For \(N=4\), the cutoffs \(k=2,3,4\) correspond respectively to
\(\bm{\sigma}_4=(0,0,0),(1,1,0),(1,1,1)\).  This cutoff family does not
include the leaf-only class \(\bm{\sigma}_4=(0,1,0)\), which is treated in
Appendix~\ref{app:analytic-qfi}.  Eqs.~\eqref{eq:cutoff-qfi}--\eqref{eq:cutoff-C}
apply to the noiseless probes; they do not provide the mixed-state QFI
after dephasing.
The \(k=N\) star-plus-all-higher-order family and the complete hypergraph
are plotted for \(N=2,\ldots,6\) in
Figs.~\ref{fig:star-all-qfi-n} and~\ref{fig:complete-qfi-n},
respectively. At \(N=2\), no higher-order hyperedges
exist, so both families have the same single edge, but their QFIs can still differ due to distinct encoding generators.

\begin{figure*}[!t]
    \centering
    \panelgraphic[width=0.49\textwidth]{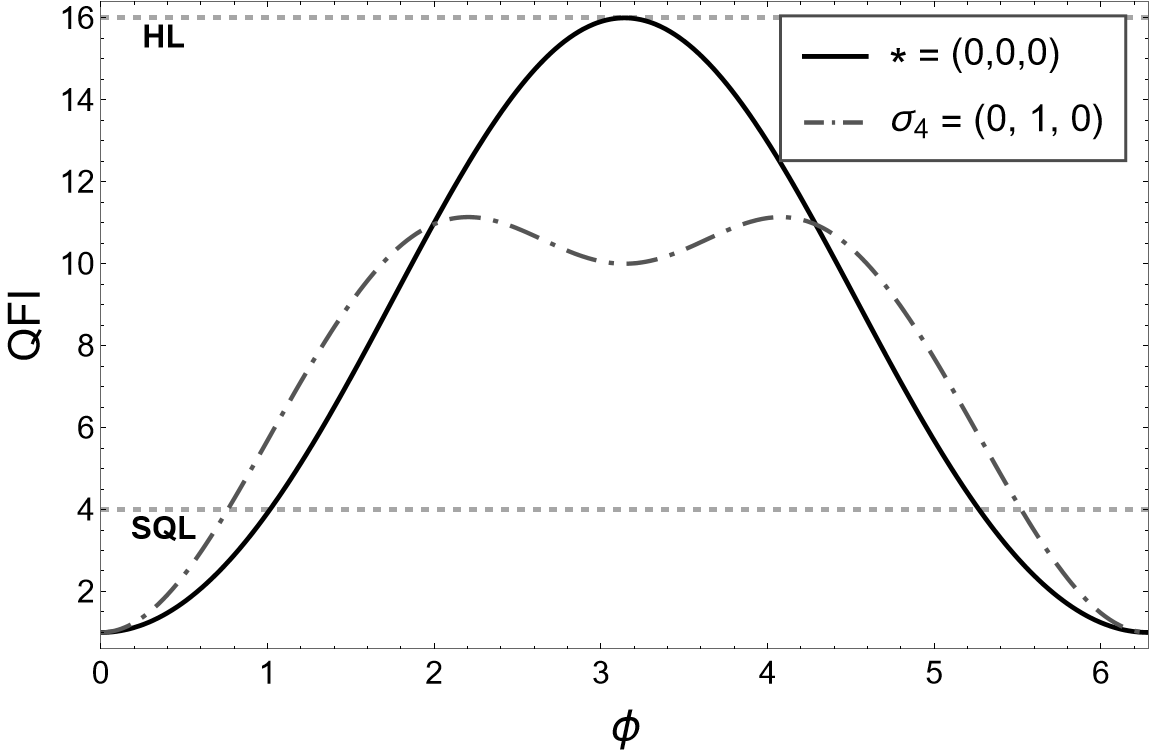}{\panellabel{12}{56}{a}}
    \hfill
    \panelgraphic[width=0.49\textwidth]{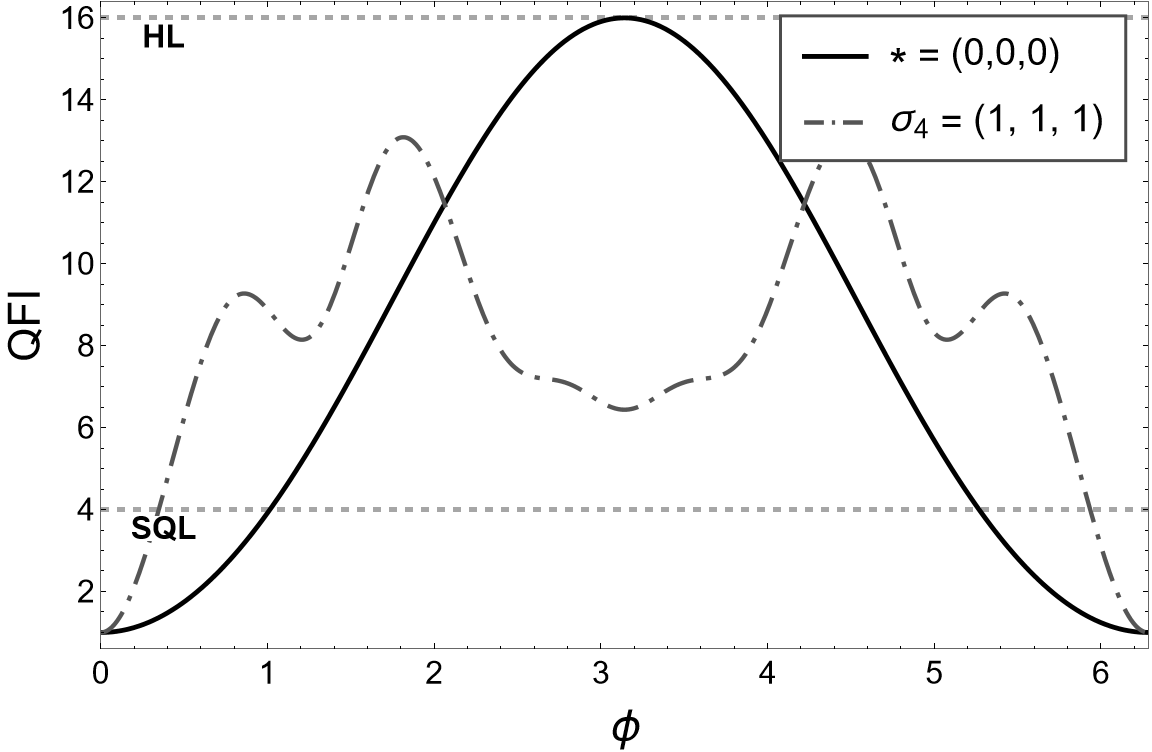}{\panellabel{12}{56}{b}}
    \caption{Representative noiseless four-qubit QFI curves.  (a) The
    class containing the single leaf-only three-qubit hyperedge,
    \(\bm{\sigma}_4=(0,1,0)\).  (b) The
    all-higher-order class \(\bm{\sigma}_4=(1,1,1)\). The gray dash-dotted
    curves represent the weighted hypergraph probes, and the black solid
    curves represent the weighted star benchmark. The horizontal dotted
    lines mark the SQL and HL.}
    \label{fig:noiseless-representatives}
\end{figure*}

\subsection{$N=4$ QFI landscape}

Fig.~\ref{fig:symmetric-landscape} compares all eight
leaf-symmetric hypergraph classes built on the star backbone of
Eq.~\eqref{eq:star-backbone} for \(0\leq\phi<2\pi\). Adding
higher-order hyperedge orbits generally introduces additional extrema and
transfers high-QFI regions away from the star optimum at \(\phi=\pi\). Several
classes consequently cross the SQL at smaller interaction weights than
the weighted star graph state, although no single class is uniformly optimal over the
entire interval. The complete-hypergraph reference assumes
Eq.~(\ref{eq:complete-encoding-generator}), so its comparison with the star-based curves
also reflects the choice of encoding generator.

Here, we focus on two representative orbit-signature classes. The probe
\(\bm{\sigma}_4=(1,1,1)\) includes every three- and four-body orbit
available at \(N=4\), whereas \(\bm{\sigma}_4=(0,1,0)\) adds only the
leaf-only orbit \(B_3\), which at \(N=4\) consists of the single
hyperedge \(\{2,3,4\}\).  As shown in Fig.~\ref{fig:noiseless-representatives}, the
leaf-only hyperedge creates a broad double-peaked profile, while the all-higher-order probe produces several distinct intervals of high QFI. The single-full-leaf family adds only one higher-order hyperedge to
the star backbone and may therefore be simpler to prepare than the
all-higher-order family, depending on the cost of implementing that
\((N-1)\)-body interaction.

On the representative interval \(0\leq\phi\leq\pi\), let
\(\phi_{\mathrm{SQL}}\) denote, separately for each class, its first upward
crossing of the SQL, and let \(\phi_{\star}\) denote its subsequent crossing
with the weighted star graph state curve.  For both classes,
\(\mathcal{F}_{\bm{\sigma}_4}(\phi,0)>
\mathcal{F}_{\star}(4,\phi,0)\) throughout
\(\phi_{\mathrm{SQL}}\leq\phi<\phi_{\star}\).  The two QFIs coincide at
\(\phi=\phi_{\star}\), after which the weighted star graph state is superior:
\(\mathcal{F}_{\star}(4,\phi,0)>
\mathcal{F}_{\bm{\sigma}_4}(\phi,0)\) for
\(\phi_{\star}<\phi\leq\pi\).  Thus the higher-order correlations shift weighted
star-beating performance to reduced interaction weights rather than improving
the weighted star graph state maximum at \(\phi=\pi\).
Numerically, the first SQL crossings are
\(\phi_{\mathrm{SQL}}\simeq0.34805\) for \((1,1,1)\) and
\(0.76158\) for \((0,1,0)\), compared with \(1.01471\) for the weighted
star graph state. The subsequent star crossings occur at
\(\phi_{\star}\simeq2.06266\) and \(1.99396\), respectively; all weights
are in radians. The ability of the star hypergraph to achieve a QFI above the SQL at a lower weight is not limited to $N=4$. We have observed similar behavior for larger $N$ using the analytical QFI expression in Eq.~\eqref{eq:cutoff-qfi} with \(k=N\) and the correlators in Eqs.~\eqref{eq:cutoff-A}--\eqref{eq:cutoff-C}.

\begin{figure*}[!t]
    \centering
    \panelgraphic[width=0.7\textwidth]{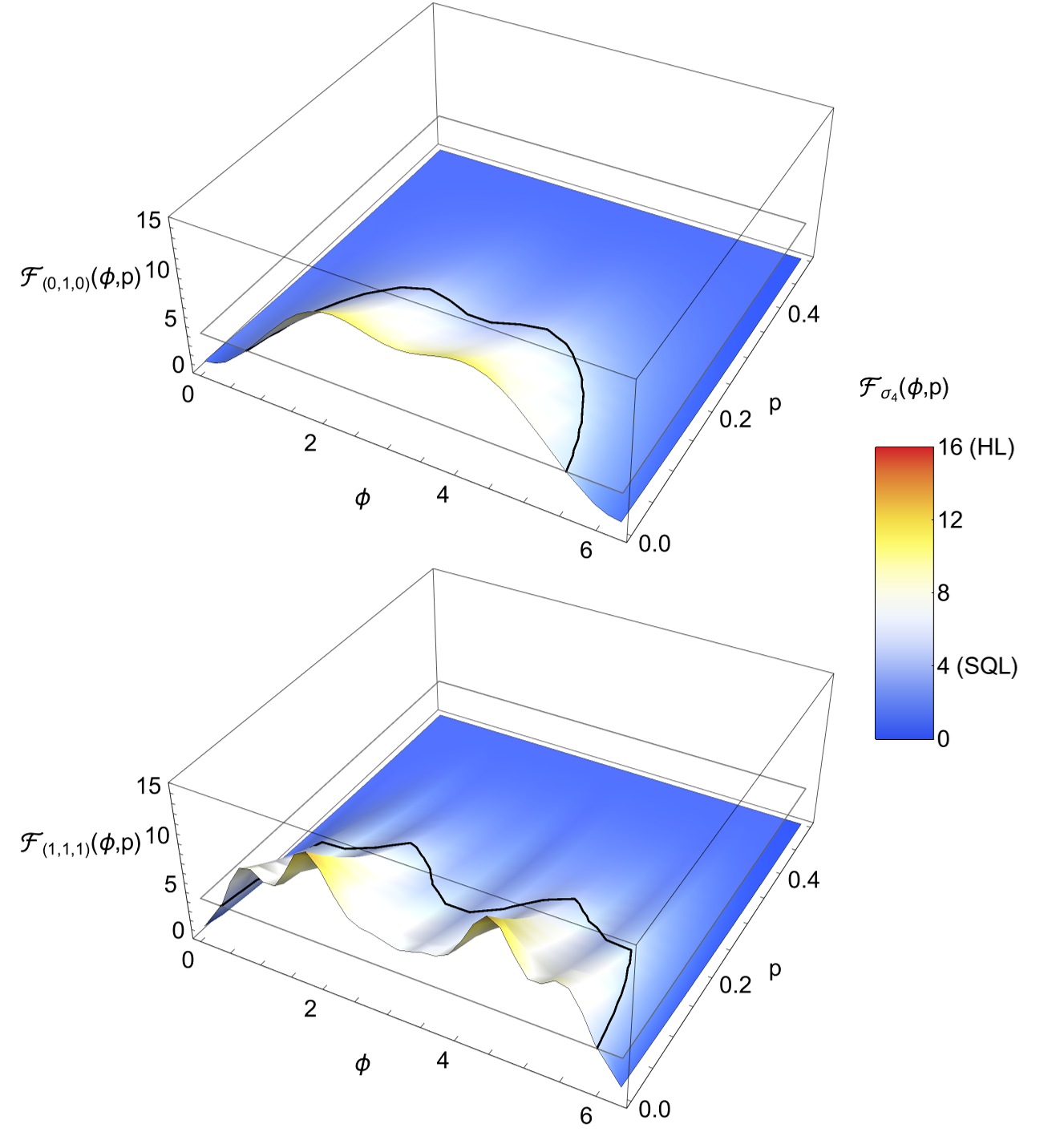}{\panellabel{2}{105}{a}\panellabel{2}{50}{b}}
    \caption{The QFI under independent local phase-flip dephasing for two
four-qubit classes. (a) \(\bm{\sigma}_4=(0,1,0)\).
(b) \(\bm{\sigma}_4=(1,1,1)\).  Surface height and color both encode
the QFI.  The horizontal plane represents the SQL at \(\mathcal{F}_Q=4\), and the
black curves trace its intersections with each surface.}
    \label{fig:dephased-qfi}
\end{figure*}

\begin{figure*}[!t]
    \centering
\panelgraphic[width=0.77\textwidth]{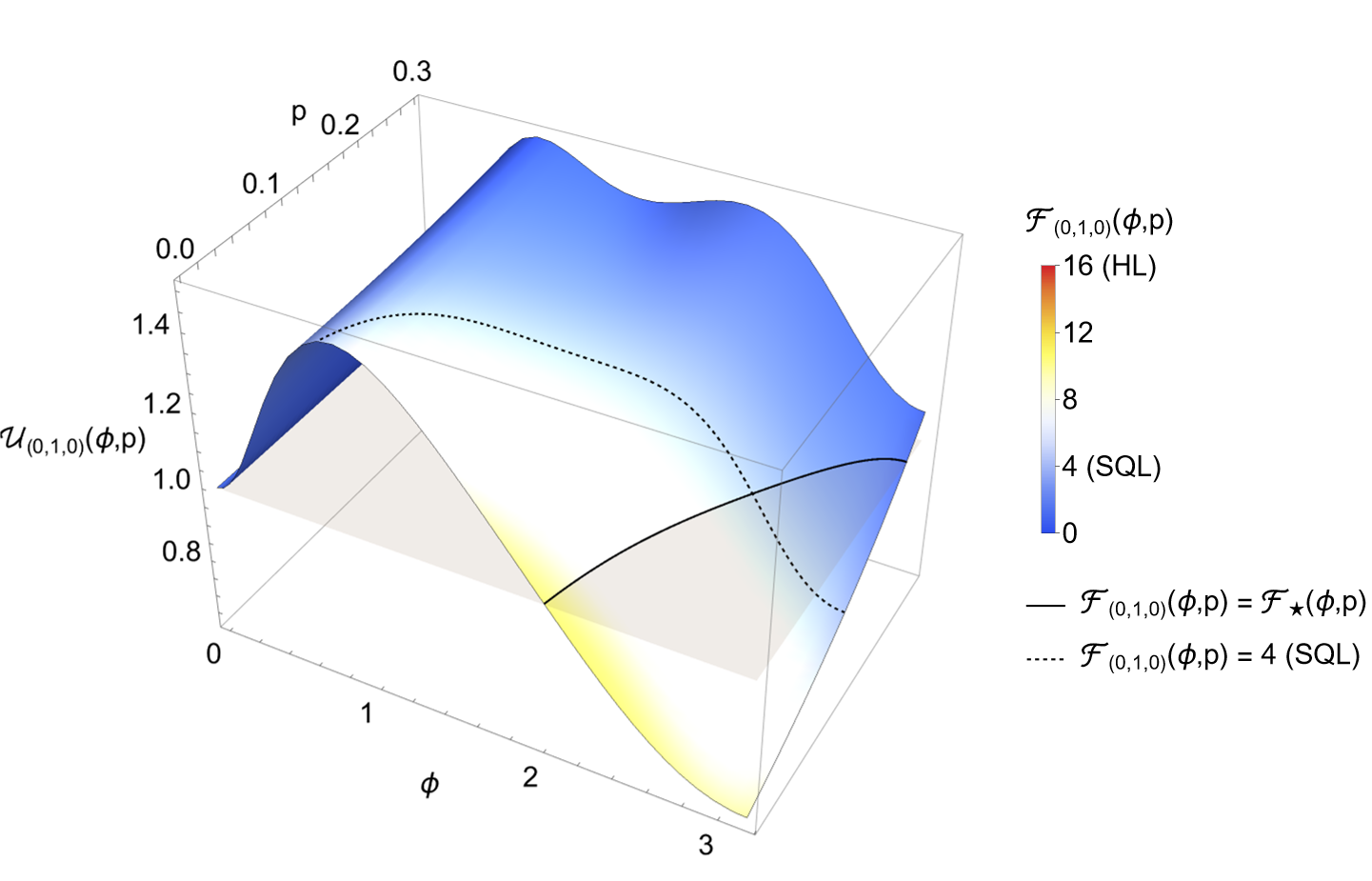}{\panellabel{2}{64.65}{a}}\\[-0.9em]
    \panelgraphic[width=0.77\textwidth]{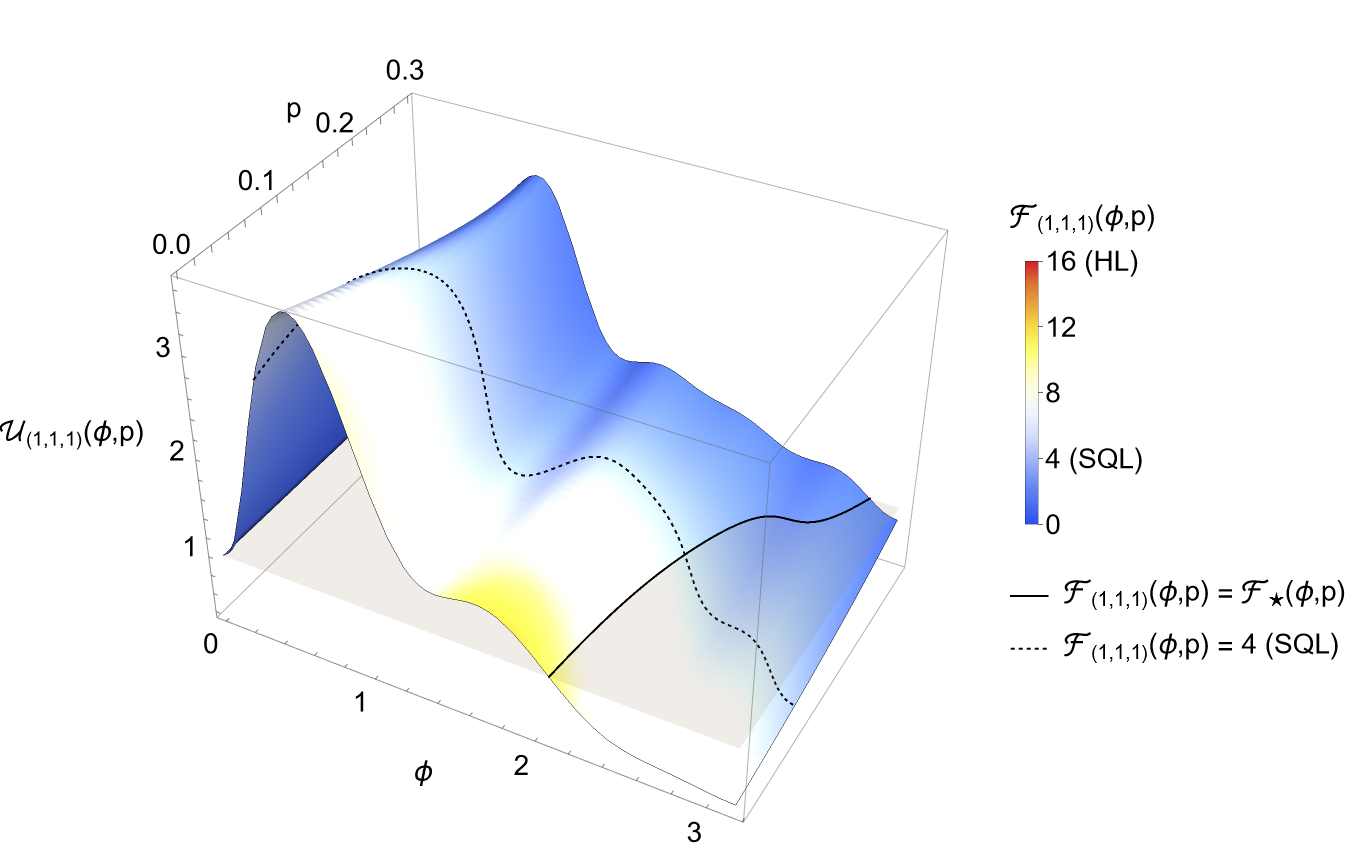}{\panellabel{2}{64.65}{b}}
    \caption{The noisy weighted star utility
    \(\mathcal{U}_{\bm{\sigma}_4}(\phi,p)\) for two four-qubit classes.
    (a) \(\bm{\sigma}_4=(0,1,0)\).
    (b) \(\bm{\sigma}_4=(1,1,1)\).  Surface height is the utility ratio,
    and surface color is the absolute hypergraph state QFI. The solid black
    curve is \(\mathcal{U}=1\), and the dotted black curve marks the SQL
    crossing \(\mathcal{F}_{\bm{\sigma}_4}=4\).}
    \label{fig:star-utility}
\end{figure*}

\begin{figure*}[!t]
    \centering
    \includegraphics[width=0.86\textwidth]{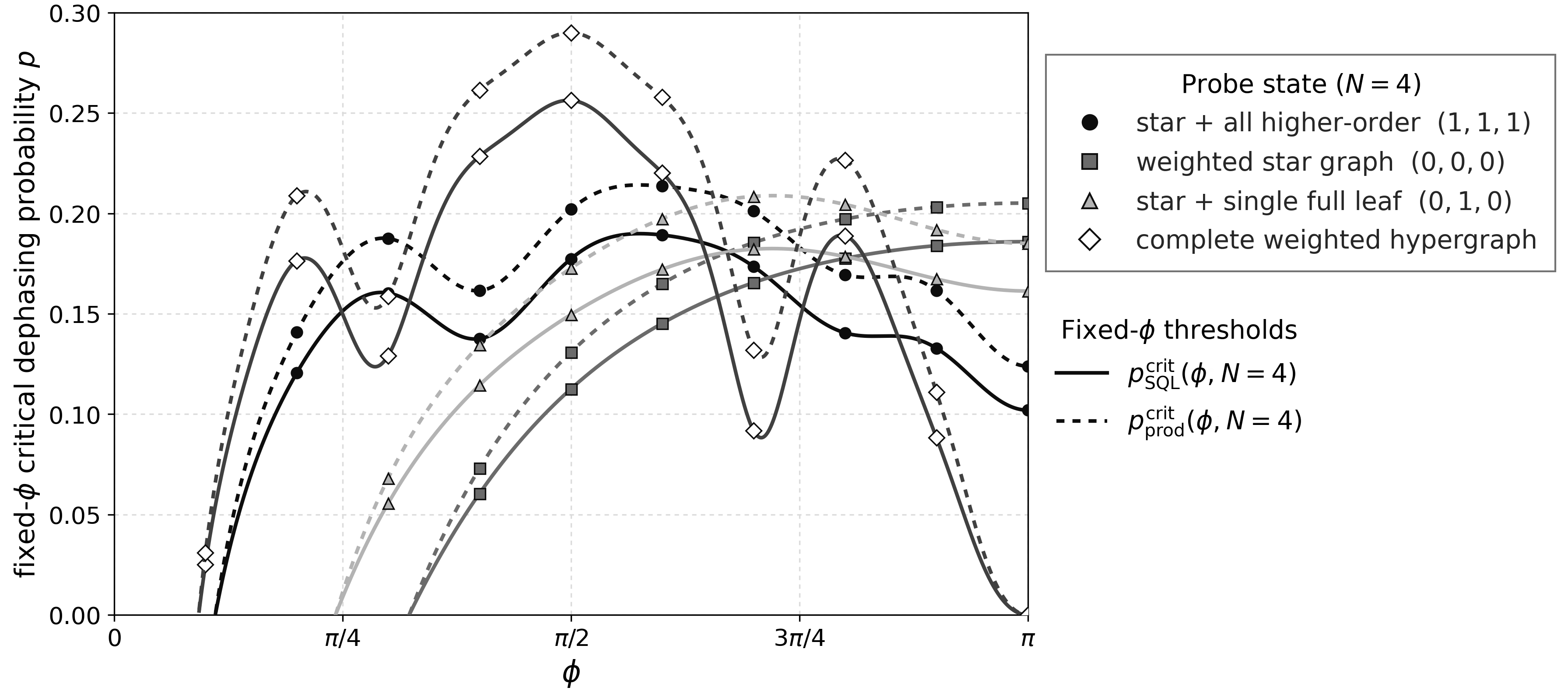}
    \caption{Fixed-\(\phi\) critical dephasing probabilities at \(N=4\).
    Black circles, dark-gray squares, and light-gray triangles identify the
    star-based orbit-signature classes \(\bm{\sigma}_4=(1,1,1)\),
    \((0,0,0)\), and \((0,1,0)\), respectively; open diamonds identify the
    complete weighted hypergraph. Solid curves give the SQL threshold
    \(\mathcal{F}_{Q}=4\), and dashed curves use the shared dephased product-state
    reference in Eq.~\eqref{eq:product-qfi}. The complete hypergraph uses
    \(\hat{H}_{\mathrm{comp}}\), for which the shared product-state reference is
    not the optimal separable bound; its dashed curve is therefore a
    shared-reference comparison. Each threshold is the upper endpoint of
    the benchmark-satisfying interval connected to \(p=0\) at fixed
    \(\phi\). Missing curve segments indicate that the benchmark is not
    met at \(p=0\); no optimization over \(\phi\) is performed.}
    \label{fig:critical-n4}
\end{figure*}

\subsection{Local dephasing}

Independent local dephasing is modeled as a $\hat{Z}$-flip channel. The single-qubit Kraus operators are given as
\begin{equation}
    \hat{K}_0=\sqrt{1-p}\,\hat{\mathbb I},
    \qquad
    \hat{K}_1=\sqrt{p}\,\hat{Z},
    \qquad
    0\leq p\leq\frac{1}{2}.
    \label{eq:dephasing-kraus}
\end{equation}
The resulting single-qubit dephasing channel is
\begin{equation}
    \mathcal{E}_{p}(\rho)
    =(1-p)\rho+p\hat{Z}\rho\hat{Z}.
    \label{eq:single-dephasing-channel}
\end{equation}
It preserves populations in the computational basis and multiplies
each off-diagonal density-matrix element by \(1-2p\). Dephasing is equivalent to phase damping using the reparameterization $p = \frac{1}{2}(1-e^{-\Gamma t})$, where $\Gamma$ is the characteristic damping rate~\cite{nielsen2002quantum}. The independent \(N\)-qubit dephasing channel is given by
\begin{align}
    \mathcal{E}_{p}^{\otimes N}(\rho)
    &=
    \sum_{\bm{m}\in\{0,1\}^{N}}
    p^{|\bm{m}|}(1-p)^{N-|\bm{m}|}
    \hat{Z}_{\bm{m}}\rho\hat{Z}_{\bm{m}},
    \label{depchan}
    % \label{eq:nqubit-dephasing-channel}
\end{align}
where $\hat{Z}_{\bm{m}} = \bigotimes_{j=1}^{N}\hat{Z}^{m_j}$ and \(|\bm m|=\sum_{j=1}^{N}m_j\) is the Hamming weight of the error string \(\bm m\).
In the noise model used here, the phase-flip channel acts on the prepared probe
before the unknown phase $\theta$ is encoded. For example, the four-qubit dephased state is
\begin{equation}
    \rho_{\bm{\sigma}_4}(\phi,p)
    =
    \mathcal{E}_{p}^{\otimes4}
    \!\left(
        \ket{\psi_{\bm{\sigma}_4}(\phi)}
        \!\bra{\psi_{\bm{\sigma}_4}(\phi)}
    \right).
\end{equation}
The parameter-dependent state entering the QFI is therefore
\begin{equation}
    \rho_{\bm{\sigma}_4,\theta}(\phi,p)
    =
    \hat{U}_{\theta}\rho_{\bm{\sigma}_4}(\phi,p)
    \hat{U}_{\theta}^{\dagger}.
    \label{eq:noisy-encoded-state}
\end{equation}
Because the preparation gates and phase-flip operators are diagonal in
the computational basis, phase-flip errors occurring during preparation can be represented as acting after preparation without changing the resulting state. 

For numerical evaluation, we write the mixed hypergraph state before encoding as
\(\rho_{\bm{\sigma}_4}(\phi,p)=\sum_a\lambda_a\ket{a}\!\bra{a}\),
where \(\lambda_a\) and \(\ket{a}\) are its eigenvalues and orthonormal eigenvectors, respectively.
For the unitary encoding in Eq.~\eqref{eq:noisy-encoded-state},
the QFI is independent of \(\theta\). The spectral
formula~\cite{liu2020quantum} then gives
\begin{equation}
    \mathcal{F}_{\bm{\sigma}_4}(\phi,p)
    =
    2\!\sum_{a,b:\lambda_a+\lambda_b>0}
    \frac{(\lambda_a-\lambda_b)^2}{\lambda_a+\lambda_b}
    \left|\bra{a}\hat{H}_{\mathrm{enc}}\ket{b}\right|^2.
    \label{eq:mixed-qfi-spectral}
\end{equation}
 For the complete-hypergraph reference, we apply the same dephasing
channel and evaluate the QFI using Eq.~\eqref{eq:mixed-qfi-spectral},
with \(\hat{H}_{\mathrm{comp}}\) in place of \(\hat{H}_{\mathrm{enc}}\).
% At \(p=1/2\), complete local dephasing erases all computational-basis
% coherences, leaving the weighted hypergraph probes maximally mixed with
% zero QFI.

The QFI surfaces in Fig.~\ref{fig:dephased-qfi} show how the useful
weight intervals contract as \(p\) increases. The \(\bm{\sigma}_4=(0,1,0)\) class remains above the SQL in one
connected region around its two noiseless lobes, while the
all-higher-order class \(\bm{\sigma}_4=(1,1,1)\) remains above it in several separate
weight intervals.

To compare a noisy hypergraph probe directly with the noisy weighted star graph state,
we define the relative metrological utility
\begin{equation}
\mathcal{U}_{\bm{\sigma}_4}(\phi,p)
=
\frac{\mathcal{F}_{\bm{\sigma}_4}(\phi,p)}
{\mathcal{F}_{\star}(4,\phi,p)}.
\label{eq:star-utility}
\end{equation}
Whenever \(\mathcal{F}_{\star}(4,\phi,p)>0\), \(\mathcal{U}>1\) identifies a QFI
improvement over the noisy star benchmark at the same \((\phi,p)\), while
\(\mathcal{U}=1\) is the crossing boundary. For comparison with the
separable reference in Eq.~\eqref{eq:product-reference}, the dephased
four-qubit product-state QFI is
\begin{equation}
\mathcal{F}_{\mathrm{prod}}(4,p)=3+(1-2p)^2.
\label{eq:product-qfi-fourqubit}
\end{equation}
For \(N\) qubits, the corresponding expression is
\begin{equation}
\mathcal{F}_{\mathrm{prod}}(N,p)
=
N-1+(1-2p)^2.
\label{eq:product-qfi}
\end{equation}
For this noise-before-encoding model, Eq.~\eqref{eq:product-qfi} is the
pointwise maximum over all fully separable input probes subjected to the same
channel, with the input state in Eq.~\eqref{eq:product-reference}
remaining optimal for all \(p\in[0,1/2]\).
Appendix~\ref{app:additional-utility}
derives this result using QFI convexity, product additivity, and the
single-qubit Bloch-vector formula~\cite{toth2014quantum,liu2020quantum}.

We analogously define the utility relative to the optimal separable benchmark,
\begin{equation}
    \mathcal{U}_{\mathrm{prod},\bm{\sigma}_4}(\phi,p)
    =
    \frac{\mathcal{F}_{\bm{\sigma}_4}(\phi,p)}
         {\mathcal{F}_{\mathrm{prod}}(4,p)},
    \label{eq:product-utility}
\end{equation}
where \(\mathcal{U}_{\mathrm{prod},\boldsymbol{\sigma}_4}>1\) indicates a QFI advantage.

Fig.~\ref{fig:star-utility} shows the metrological utility of the two representative four-qubit hypergraph probes relative to the equally dephased weighted star graph state, isolating relative performance from the overall decay of QFI.  The unity plane separates parameter pairs where the
hypergraph state improves on the star from those where it does not.  The
all-higher-order probe exhibits the larger and more structured relative gains,
whereas the leaf-only hyperedge provides a more modest advantage over a
restricted region.  A large utility ratio is metrologically relevant only when the corresponding QFI, shown by the color scale, also exceeds the desired benchmark.  Product-state utility plots are presented in Appendix~\ref{app:additional-utility}.

\subsection{Critical dephasing probabilities}
We assess dephasing tolerance against both the ideal noiseless SQL,
\(\mathcal{F}_Q=N\), and the equally dephased product-state reference
in Eq.~\eqref{eq:product-qfi}, the latter providing a fairer
experimental comparison under matched noise conditions. Define the benchmarks
\(B_{\mathrm{SQL}}(N,p)=N\) and
\(B_{\mathrm{prod}}(N,p)=\mathcal{F}_{\mathrm{prod}}(N,p)\). At fixed \(\phi\), we define the critical probability for a probe family \(\mathcal{G}\) as the upper
endpoint of the benchmark-satisfying interval connected to \(p=0\),
\begin{equation}
\begin{split}
p_{B,\mathcal{G}}^{\mathrm{crit}}(\phi,N)
=\sup\!\bigl\{p\in[0,1/2]:{}&
\mathcal{F}_{\mathcal{G}}(N,\phi,p')\geq B(N,p')\\[-2pt]
&\text{for every }p'\in[0,p]\bigr\},
\end{split}
\label{eq:pointwise-critical-probabilities}
\end{equation}
where \(B\in\{B_{\mathrm{SQL}},B_{\mathrm{prod}}\}\). In other words, at a fixed preparation weight, this threshold gives the maximum dephasing probability the probe can tolerate before its QFI first falls below the chosen benchmark as noise increases from zero. The threshold is
left undefined when the probe does not meet the benchmark at \(p=0\). The shared product-state reference in Eq.~\eqref{eq:product-qfi}
is optimal for separable probes under \(\hat{H}_{\star}\);
for the complete hypergraph, it is a shared comparison
reference rather than the optimal separable bound under
\(\hat{H}_{\mathrm{comp}}\).

Fig.~\ref{fig:critical-n4} plots the fixed-\(\phi\) critical
dephasing probabilities at \(N=4\).
The solid triangle and circle curves give the SQL-crossing thresholds
for the leaf-only and all-higher-order probe states shown in
Fig.~\ref{fig:dephased-qfi}(a) and (b), respectively.
The dashed curves use the equally dephased product-state reference
and lie at or above the corresponding solid curves for all families,
since \(\mathcal{F}_{\mathrm{prod}}(4,p)\leq4\).
The thresholds vary appreciably with the interaction weight~\(\phi\).
Among the three star-based families, the all-higher-order class
\(\bm{\sigma}_4=(1,1,1)\) has the largest thresholds over much of the
intermediate-weight region, with its maxima shifted away from
\(\phi=\pi\).
The complete-hypergraph reference reaches still higher thresholds
around \(\phi=\pi/2\), but uses the distinct generator
\(\hat{H}_{\mathrm{comp}}\).
The leaf-only class \(\bm{\sigma}_4=(0,1,0)\) also shifts the useful
region to lower weights, although its ordering relative to the star
depends on the weight and benchmark.
Near \(\phi=\pi\), the weighted star has the largest displayed
thresholds.
No probe is uniformly superior, but selected hypergraphs improve dephasing tolerance over specific weight intervals below \(\phi=\pi\).

\begin{figure*}[!t]
    \centering
    \panelgraphic[width=1.\textwidth]{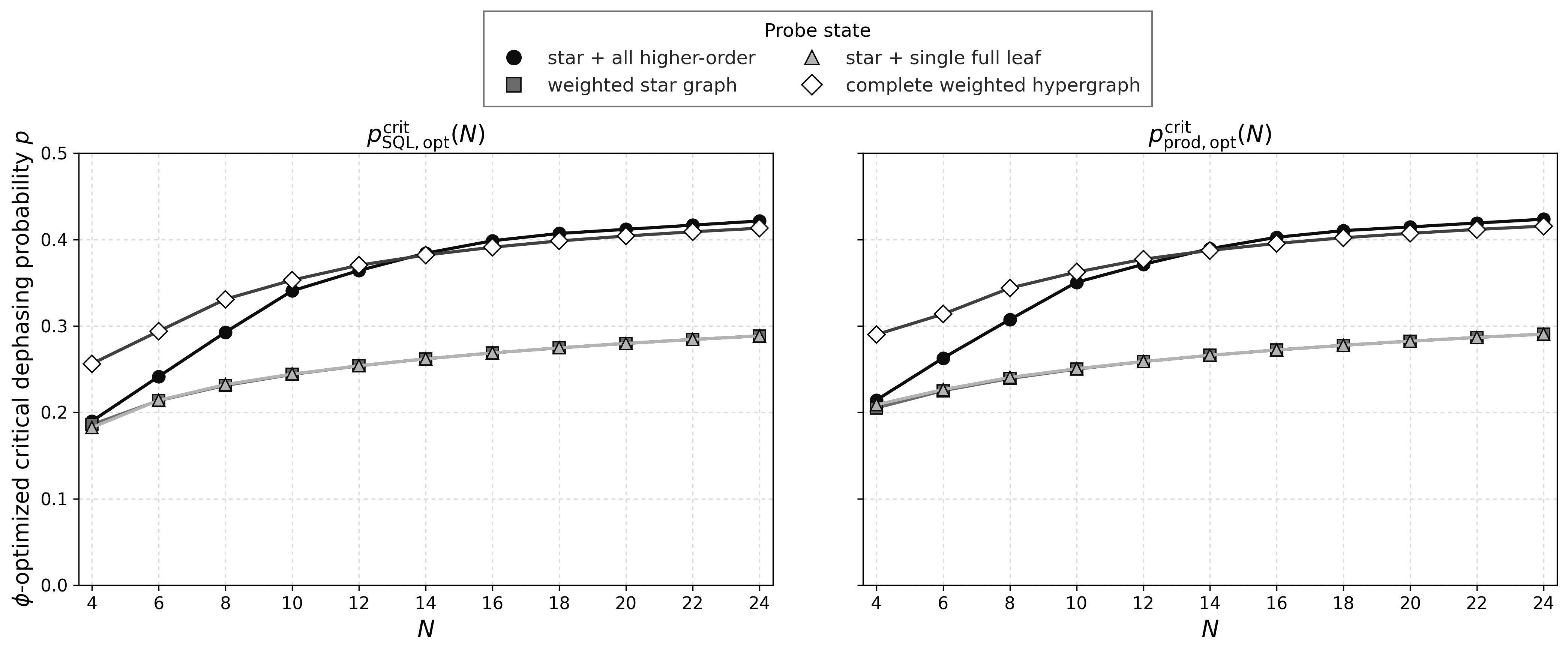}{\panellabel{6}{30}{a}\panellabel{56}{30}{b}}
    \caption{Numerically \(\phi\)-optimized critical dephasing
    probabilities for even \(N=4,6,\ldots,24\).
    (a) Thresholds against the SQL of \(\mathcal F_Q=N\).
    (b) Thresholds against the shared product-state reference of
    Eq.~\eqref{eq:product-qfi}, which is weaker than the complete
    family's optimal separable bound.
    Black circles, dark-gray squares, and light-gray triangles
    denote the all-higher-order, weighted star, and
    single-full-leaf families, whose four-qubit orbit signatures are
    \((1,1,1)\), \((0,0,0)\), and \((0,1,0)\), respectively.
    Open diamonds denote the complete weighted hypergraph.
    The three star-based families use \(\hat H_{\star}\); the
    complete family uses \(\hat H_{\mathrm{comp}}\).
    The two thresholds are optimized separately
    over \(\phi\) for each family and \(N\).}
    \label{fig:critical-scaling}
\end{figure*}

The family-level threshold used in the scaling comparison is
\begin{equation}
    p_{B,\mathcal{G},\mathrm{opt}}^{\mathrm{crit}}(N)
    =
    \max_{0\leq\phi\leq\pi}
    p_{B,\mathcal{G}}^{\mathrm{crit}}(\phi,N).
    \label{eq:optimized-critical-probabilities}
\end{equation}
Eqs.~\eqref{eq:pointwise-critical-probabilities} and
\eqref{eq:optimized-critical-probabilities} distinguish the threshold at a
prescribed interaction weight from the best threshold available if one is able to
retune that weight.  The SQL and product-state thresholds are optimized
separately and therefore need not select the same value of \(\phi\).
The maximization is taken only over weights for which the corresponding
fixed-weight threshold is defined.

Appendix~\ref{app:additional-utility}
presents product-state utility plots and fixed-\(p\) sections for the classes $\boldsymbol{\sigma}_4 = (0,1,0)$ and $(1,1,1)$.
For the optimized dephasing thresholds at even \(N=4,6,\ldots,24\),
Appendix~\ref{app:critical-scaling} gives the family definitions and
numerical methods.

In Fig.~\ref{fig:critical-scaling}, all eight threshold sequences increase with $N$, and the weighted star optimum is \(\phi=\pi\)  for both references. The single-full-leaf thresholds closely track the weighted star, while the all-higher-order family has the largest thresholds among the star-based families. In both panels, the complete family leads through $N = 12$, whereas the all-higher-order family leads at the sampled sizes $N = 14$ to $N = 24$. At $N = 24$, their SQL thresholds are approximately $0.421410$ for the all-higher-order family and $0.413111$ for the complete family. The shared product-state reference is less demanding than the SQL, so probes can tolerate at least as much dephasing before falling below it. For the complete-family generator, however, the state \(\ket{0}^{\otimes N}\) is unaffected by dephasing and attains a QFI of $N$, the optimal separable bound given in Eq.~\eqref{eq:complete-separable-bound}. The threshold against this bound therefore equals the SQL threshold, while panel (b) uses the weaker shared reference. The complete and star-based families use different generators, so their comparison concerns probe–generator combinations.

\section{Discussion}
In this work, we studied weighted hypergraph states as probes for quantum metrology, asking how higher-order interaction patterns and preparation weights affect phase sensitivity and robustness to local dephasing. We combined analytical quantum Fisher information expressions and a symmetry-adapted Dicke-state representation with numerical calculations, focusing on four-qubit probes and selected families at larger system sizes. Our aim was to identify hypergraph structures that support a metrological advantage over separable and weighted graph-state benchmarks, and to assess how this advantage persists under noise.

Our results show that higher-order interactions can yield weighted
hypergraph probe states that surpass the SQL and outperform the weighted
star graph state over identified ranges of weights and dephasing probabilities. Equally dephased product-state benchmarks complement the
SQL by quantifying gains over separable references under the same
noise. Complete higher-order hyperedge orbits can shift enhanced
phase sensitivity to smaller interaction weights than those required
by the weighted star graph state, a trend observed for four qubits and at
larger system sizes. For $N=4$, the effect is already visible
for the leaf-only three-body hyperedge \(B_3\) and becomes stronger
when all three- and four-body orbits are included. Both representatives cross the SQL and outperform the weighted star graph state at smaller weights, although the latter performs better at \(\phi=\pi\).

The QFI surfaces show where the probes exceed the SQL, while
\(\mathcal{U}\) measures gains over the equally noisy weighted star graph state,
and the critical probabilities quantify noise tolerance relative to
each benchmark at fixed or optimized \(\phi\). The four-qubit all-higher-order
class \(\bm{\sigma}_4=(1,1,1)\) exhibits broader and more structured
gains than the leaf-only class. 

After optimization over \(\phi\), the threshold advantage of the all-higher-order star-based family over the weighted star generally increases across the displayed even sizes \(4\leq N\leq24\). Its thresholds are below the
complete-hypergraph reference at smaller sizes and slightly above it
at larger sizes. The single-full-leaf generalization of
\(\bm{\sigma}_4=(0,1,0)\) remains close to the star. Its SQL threshold
is lower at \(N=4\) and marginally higher at the other plotted sizes,
while its threshold against the shared product-state reference is slightly
higher throughout. 

This study establishes weighted hypergraph states as tunable
metrological resources whose higher-order hyperedge structure can
enhance phase sensitivity and sustain metrological advantages under
local dephasing. The derived binomial correlator formulas and the \(2N\)-dimensional
pure-state representation reduce simulation costs by allowing the pure-state
QFI to be evaluated without constructing the full
\(2^N\)-dimensional state vector.
Future work could extend the analytic results to other hypergraph state classes and non-uniform weights, optimize encoding generators, identify experimentally accessible near-optimal measurements, and quantify the costs of state preparation and measurement. The framework could also be extended to multiparameter estimation~\cite{toth2014quantum,liu2020quantum} and distributed quantum sensing~\cite{kim2025distributed}.

\section{Acknowledgments}
We acknowledge Stellenbosch University, the National Research Foundation, and the Department of Science and Innovation through the South African Quantum Technology Initiative for funding this project.

% \section*{Data and code availability}
% Numerical data, validation scripts, and production code are provided in the
% accompanying source archive under \texttt{numerical\_checks/}.

\appendix

\section{General symmetric representation}
\label{app:symmetric-representation}

For \(N\geq3\), with one central qubit and \(N-1\) interchangeable leaves,
the orbit classes in Eqs.~\eqref{eq:general-orbits} and~\eqref{eq:general-orbits2} provide \(N-2\)
central indicators \(a_r\) and \(N-3\) leaf-only indicators \(b_r\).
For \(N\geq4\), we order them as
\begin{equation}
\bm{\sigma}_N
=
(a_3,b_3,a_4,b_4,\ldots,a_{N-1},b_{N-1},a_N)
\in\{0,1\}^{2N-5},
\label{eq:general-orbit-signature}
\end{equation}
which gives \(\bm{\sigma}_4=(a_3,b_3,a_4)\) at \(N=4\). For \(N=3\),
there is only the central orbit indicator \(\bm{\sigma}_3=(a_3)\).
The phase-count function for central bit \(c\) and leaf Hamming weight \(w\) is
\begin{equation}
g_{\bm{\sigma}_N}(c,w)
=
cw
+
\sum_{r=3}^{N}a_r c\binom{w}{r-1}
+
\sum_{r=3}^{N-1}b_r\binom{w}{r}.
\label{eq:general-phase-count}
\end{equation}
Each sum is empty when its upper limit is below~3. Let
\(\ket{D_w^{N-1}}=\binom{N-1}{w}^{-1/2}
\sum_{y\in\{0,1\}^{N-1}:\,|y|=w}\ket{y}\)
be the normalized symmetric Dicke state of the leaves.
The corresponding ideal probe can be written as
\begin{equation}
\begin{aligned}
\ket{\psi_{\bm{\sigma}_N}(\phi)}
={}&
\frac{1}{\sqrt{2^N}}
\sum_{c=0}^{1}\sum_{w=0}^{N-1}
\sqrt{\binom{N-1}{w}}
\\
&\times
e^{i\phi g_{\bm{\sigma}_N}(c,w)}
\ket{c}\ket{D_w^{N-1}}.
\end{aligned}
\label{eq:general-symmetric-probe}
\end{equation}
It lies in the subspace
\begin{equation}
\mathbb{C}^{2}\otimes
\operatorname{Sym}^{N-1}(\mathbb{C}^{2}),
\qquad
\dim=2N,
\end{equation}
of the full \(2^N\)-dimensional Hilbert space. The number of binary
orbit indicators is \(2N-5\), so this construction defines
\(2^{2N-5}\) symmetric orbit selections.

Separately, the complete weighted hypergraph of
Eq.~\eqref{eq:complete-hypergraph-edges} is symmetric under permutations
of all \(N\) qubits. Its phase count depends only on the total Hamming
weight \(s\) and is \(g_{\mathrm{comp}}(s)=2^s-1-s\), as shown in
Eq.~\eqref{eq:proof-complete-phase}. Let
\(\ket{D_s^N}=\binom{N}{s}^{-1/2}
\sum_{x\in\{0,1\}^{N}:\,|x|=s}\ket{x}\)
be the normalized Dicke state of all \(N\) qubits. Grouping
computational-basis strings of equal Hamming weight gives
\begin{equation}
\begin{aligned}
\ket{\psi_{\mathrm{comp}}(\phi)}
&=2^{-N/2}\sum_{s=0}^{N}
\sqrt{\binom{N}{s}}\\
&\quad\times e^{i\phi g_{\mathrm{comp}}(s)}
\ket{D_s^N}.
\end{aligned}
\label{eq:complete-symmetric-probe}
\end{equation}
The ideal complete-hypergraph probe therefore lies in
\(\operatorname{Sym}^{N}(\mathbb{C}^{2})\), an
\((N+1)\)-dimensional subspace of the full \(2^N\)-dimensional
Hilbert space.

\section{Proof of the analytic QFI}
\label{app:analytic-qfi}

\subsection{Complete hypergraph state QFI}
\label{app:complete-qfi-proof}
For the edge set in Eq.~\eqref{eq:complete-hypergraph-edges}, a basis string
\(x\) of Hamming weight \(s=|x|\) activates \(\binom{s}{m}\)
hyperedges of order \(m\). Its phase count is therefore
\begin{equation}
    g_{\mathrm{comp}}(s)
    =\sum_{m=2}^{N}\binom{s}{m}
    =2^s-1-s,
    \label{eq:proof-complete-phase}
\end{equation}
and the uniformly weighted probe is
\begin{equation}
    \ket{\psi_{\mathrm{comp}}(\phi)}
    =2^{-N/2}\sum_{x\in\{0,1\}^N}
    e^{i\phi g_{\mathrm{comp}}(|x|)}\ket{x}.
    \label{eq:proof-complete-state}
\end{equation}
Because its phase depends only on \(s\), all qubits are permutation
equivalent. With
\(\hat{S}_{\mathrm{comp}}=2\hat{H}_{\mathrm{comp}}
=\sum_{j=1}^{N}\hat{X}_j\), define
\(\mathsf{A}_{N}^{\mathrm{comp}}=\langle\hat{X}_1\rangle\) and
\(\mathsf{C}_{N}^{\mathrm{comp}}=\langle\hat{X}_1\hat{X}_2\rangle\).
The moments are
\begin{align}
    \langle\hat{S}_{\mathrm{comp}}\rangle
    &=N\mathsf{A}_{N}^{\mathrm{comp}},\\
    \langle\hat{S}_{\mathrm{comp}}^2\rangle
    &=N+N(N-1)\mathsf{C}_{N}^{\mathrm{comp}}.
    \label{eq:proof-complete-moments}
\end{align}
Taking the variance of \(\hat{S}_{\mathrm{comp}}\) in the pure
state gives Eq.~\eqref{eq:complete-hypergraph-qfi} via Eq.~\eqref{FQH}.

To calculate \(\mathsf{A}_{N}^{\mathrm{comp}}\) in the state
\(\ket{\psi_{\mathrm{comp}}(\phi)}\), fix an assignment of the
remaining \(N-1\) qubits with Hamming weight \(r\).
The operator \(\hat X_1\) flips qubit~1, pairing
computational-basis strings of total Hamming weights
\(r\) and \(r+1\), with phase-count difference
\begin{equation}
g_{\mathrm{comp}}(r+1)-g_{\mathrm{comp}}(r)=2^r-1.
\label{eq:proof-complete-single-increment}
\end{equation}
The two terms from this pair in \(\langle\hat X_1\rangle\) have
complex-conjugate phase factors. Since each amplitude product has
magnitude \(2^{-N}\), their sum is
\begin{equation}
2^{-N}\!\left(e^{i\phi(2^r-1)}
+e^{-i\phi(2^r-1)}\right)
=2^{1-N}\cos\!\left[\phi(2^r-1)\right].
\label{eq:proof-complete-single-pair-sum}
\end{equation}
There are \(\binom{N-1}{r}\) configurations of the remaining
\(N-1\) qubits with Hamming weight \(r\). Summing their contributions
over \(r=0,\ldots,N-1\) yields Eq.~\eqref{eq:complete-hypergraph-A}.

For \(\mathsf{C}_{N}^{\mathrm{comp}}\), fix an assignment of the
remaining \(N-2\) qubits with Hamming weight \(r\). The operator
\(\hat X_1\hat X_2\) exchanges the pairs
\(01\leftrightarrow10\) and \(00\leftrightarrow11\).
The strings in the first pair both have total Hamming weight \(r+1\),
so their two terms have equal phase and sum to \(2^{1-N}\).
The strings in the second pair have phase-count difference
\begin{equation}
g_{\mathrm{comp}}(r+2)-g_{\mathrm{comp}}(r)
=3\cdot2^r-2.
\label{eq:proof-complete-pair-increment}
\end{equation}
Their two terms sum to
\(2^{1-N}\cos[\phi(3\cdot2^r-2)]\). Thus each assignment
of the remaining qubits with Hamming weight \(r\) contributes
\begin{equation}
2^{1-N}\left\{1+
\cos\!\left[\phi(3\cdot2^r-2)\right]\right\}
\label{eq:proof-complete-two-flip-contribution}
\end{equation}
to \(\mathsf{C}_{N}^{\mathrm{comp}}\). There are
\(\binom{N-2}{r}\) such assignments. Summing their contributions
over \(r=0,\ldots,N-2\) yields Eq.~\eqref{eq:complete-hypergraph-C}.
% In particular, at \(N=4\),
% \begin{align}
%     \mathsf{A}_{4}^{\mathrm{comp}}
%     &=\tfrac18\bigl(1+3\cos\phi+3\cos3\phi+\cos7\phi\bigr),\\
%     \mathsf{C}_{4}^{\mathrm{comp}}
%     &=\tfrac18\bigl(4+\cos\phi+2\cos4\phi+\cos10\phi\bigr).
%     \label{eq:proof-complete-n4-correlators}
% \end{align}
% Substitution into Eq.~\eqref{eq:complete-hypergraph-qfi} gives the
% four-qubit complete-hypergraph curve in
% Fig.~\ref{fig:symmetric-landscape}.
% As a check, at \(\phi=0\) the correlators satisfy
% \(\mathsf{A}_{N}^{\mathrm{comp}}=
% \mathsf{C}_{N}^{\mathrm{comp}}=1\), giving zero QFI because
% \(\ket{+}^{\otimes N}\) is an eigenstate of
% \(\hat H_{\mathrm{comp}}\). Direct state-vector evaluations agree with the
% cutoff formula in 2112 cases (\(N=3,\ldots,10\), every
% \(2\leq k\leq N\), and 48 weights per pair) to a maximum absolute
% difference of \(9.52\times10^{-13}\). The complete-hypergraph formula
% agrees in 384 cases (\(N=3,\ldots,10\), 48 weights each) to
% \(1.89\times10^{-12}\), in double precision. These finite checks
% supplement the derivations above. For \(N=2\), the star-plus-all-higher-order
% curve in Fig.~\ref{fig:star-all-qfi-n} follows directly from
% Eq.~\eqref{eq:star-qfi}; the cutoff correlator sums require \(N\geq3\).
\subsection{Star-based cutoff-family QFI}
\label{app:cutoff-qfi-proof}
We derive the finite-sum expressions for
\(\mathsf{A}_{N,k}\), \(\mathsf{B}_{N,k}\), and
\(\mathsf{C}_{N,k}\) in
Eqs.~\eqref{eq:cutoff-A}--\eqref{eq:cutoff-C}.
These correlators determine the pure-state QFI through
Eq.~\eqref{eq:cutoff-qfi} for the cutoff edge set
\(E_{N,k}\) defined in Eq.~\eqref{eq:cutoff-edge-set}. Write a computational-basis string as
\(x=(c,x_2,\ldots,x_N)\), where \(c=x_1\in\{0,1\}\) is the central bit,
\(x_j\in\{0,1\}\) are the leaf bits, and
\(w=\sum_{j=2}^{N}x_j\) is the leaf Hamming weight.
For this basis string, exactly \(cw\) star pair-edges
\(\{1,j\}\) have both endpoint bits equal to one:
when \(c=0\), there are none; when \(c=1\), there is
one for each of the \(w\) leaves with \(x_j=1\).
Since \(E_{N,k}\) contains every \(m\)-qubit hyperedge for
\(m=3,\ldots,k\), exactly \(\binom{c+w}{m}\) of these hyperedges
have all their vertex bits equal to one.
For any included star pair-edge or higher-order hyperedge
\(e\subseteq V\), its phase gate multiplies the
computational-basis component \(\ket{x}\) by \(e^{i\phi}\)
when \(x_j=1\) for every \(j\in e\), and by unity otherwise. For example, \(N=4\), \(k=3\), and \(x=(1,1,0,1)\)
give \(c=1\) and \(w=2\). Thus \(cw=2\) star edges and
\(\binom{c+w}{3}=1\) three-qubit hyperedge contribute,
giving the phase factor \(e^{3i\phi}\).

Extending the phase-count notation of Eq.~\eqref{eq:n4-phase-count}
to the cutoff family \(E_{N,k}\), we define \(g_k(c,w)\)
as the number of included star pair-edges
and higher-order hyperedges whose vertices all have bit value one
in the computational-basis string \(x\). Let \(T_k(s)\) count the
higher-order hyperedges with this property when \(x\) has total
Hamming weight \(s=|x|=c+w\). Then
\begin{align}
    g_k(c,w)&=cw+T_k(c+w),
    \label{eq:proof-phase-count}\\
    T_k(s)&=\sum_{m=3}^{k}\binom{s}{m}.
    \label{eq:proof-T}
\end{align}
The commuting diagonal gates therefore prepare
\begin{equation}
    \ket{\psi_{N,k}(\phi)}
    =
    2^{-N/2}\sum_{x\in\{0,1\}^{N}}
    e^{i\phi g_k(x)}\ket{x},
    \label{eq:proof-state}
\end{equation}
where \(g_k(x)\equiv g_k(c,w)\).  The amplitudes have equal magnitude, so
\(\langle\hat{Z}_1\rangle=0\), and the phase count is invariant under all
leaf permutations.

\paragraph*{Reduction to three correlators.} We now use leaf-permutation symmetry to express the pure-state QFI in terms of the three correlators given by Eqs.~\eqref{eq:cutoff-A}--\eqref{eq:cutoff-C}.
Let \(\hat S=2\hat H_{\mathrm{enc}}\). For the pure probe,
\(\mathcal F_Q=4\,\operatorname{Var}(\hat H_{\mathrm{enc}})
=\operatorname{Var}(\hat S)\), so we compute the first two
moments of \(\hat S\).  Since the \(N\) terms in
\(\hat{S}=\hat{Z}_1+\sum_{j=2}^{N}\hat{X}_j\) act on distinct qubits,
\begin{equation}
    \hat{S}^{2}
    =
    N\hat{\mathbb I}
    +2\sum_{j=2}^{N}\hat{Z}_1\hat{X}_j
    +2\sum_{2\leq j<\ell\leq N}\hat{X}_j\hat{X}_\ell.
    \label{eq:proof-S-square}
\end{equation}
There are \(N-1\) central--leaf terms and
\(\binom{N-1}{2}\) leaf--leaf terms.  Leaf symmetry consequently gives
\begin{align}
    \langle\hat{S}\rangle
    &=(N-1)\mathsf{A}_{N,k},\\
    \langle\hat{S}^{2}\rangle
    &=N+2(N-1)\mathsf{B}_{N,k}\nonumber\\
    &\quad +(N-1)(N-2)\mathsf{C}_{N,k}.
    \label{eq:proof-S-moments}
\end{align}
Eq.~\eqref{eq:cutoff-qfi} follows from 
\(\mathcal{F}_{Q}=\operatorname{Var}(\hat{S}) = \langle \hat{S}^2 \rangle -\langle \hat{S}\rangle^2 \).

\paragraph*{Phase differences under leaf flips.}
For \(R\subseteq L\), let \(x\oplus R\) denote the string with all bits
in \(R\) flipped, and set \(\hat{X}_R=\prod_{j\in R}\hat{X}_j\).
Since $R \subseteq L$, the central bit $c$ remains unchanged.  For $\eta \in \{0,1\}$,
Eq.~\eqref{eq:proof-state} gives the identity
\begin{equation}
    \begin{aligned}
        \bigl\langle\hat{Z}_1^\eta\hat{X}_R\bigr\rangle
        ={}&2^{-N}\sum_{x\in\{0,1\}^{N}}(-1)^{\eta c}\\
        &\times e^{i\phi[g_k(x\oplus R)-g_k(x)]}.
    \end{aligned}
    \label{eq:proof-flip-identity}
\end{equation}
Pascal's identity yields
\begin{equation}
    \begin{aligned}
        T_k(s+1)-T_k(s)
        &=
        \sum_{m=3}^{k}
        \left[\binom{s+1}{m}-\binom{s}{m}\right]\\
        &=\sum_{q=2}^{k-1}\binom{s}{q}=D_k(s),
    \end{aligned}
    \label{eq:proof-finite-difference}
\end{equation}
while a change of two in the total Hamming weight gives
\(T_k(s+2)-T_k(s)=D_k(s)+D_k(s+1)\).

For a single-leaf flip, choose a leaf \(j\in L\) and hold
the central bit \(c\) and all other leaf bits fixed. Let $r$
be the Hamming weight of these \(N-2\) other leaves. The operator
\(\hat X_j\) pairs the strings with \(x_j=0\) and \(x_j=1\),
whose leaf weights are \(r\) and \(r+1\), respectively.
Their phase-count difference is
\begin{equation}
    \begin{aligned}
        g_k(c,r+1)-g_k(c,r)
        &=c+\bigl[T_k(c+r+1)-T_k(c+r)\bigr]\\
        &=c+D_k(c+r)
        =\Delta_k^{(1)}(c,r).
    \end{aligned}
    \label{eq:proof-single-increment}
\end{equation}
For the string with \(x_j=0\), its term in
Eq.~\eqref{eq:proof-flip-identity} is
\((-1)^{\eta c}e^{i\phi\Delta_k^{(1)}(c,r)}\).
For the otherwise identical string with \(x_j=1\), the leaf
flip reverses the phase-count difference, giving
\((-1)^{\eta c}e^{-i\phi\Delta_k^{(1)}(c,r)}\).
The central bit is unchanged, so these two terms sum to
\(2(-1)^{\eta c}\cos[\phi\Delta_k^{(1)}(c,r)]\).
There are \(\binom{N-2}{r}\) configurations of the other leaves.
Setting \(\eta=0\), summing over \(c,r\), and multiplying by
\(2^{-N}\) gives Eq.~\eqref{eq:cutoff-A}.
Taking \(\eta=1\) supplies the additional factor \((-1)^c\)
and gives Eq.~\eqref{eq:cutoff-B}.

For a two-leaf flip, \(r\) instead counts occupied qubits among the
remaining \(N-3\) leaves.  The configurations \(00\) and \(11\) are
paired, with endpoint increment
\begin{equation}
    \begin{aligned}
        &g_k(c,r+2)-g_k(c,r)\\
        &\quad=2c+\bigl[T_k(c+r+2)-T_k(c+r)\bigr]\\
        &\quad=2c+\bigl[T_k(c+r+1)-T_k(c+r)\bigr]\\
        &\qquad+\bigl[T_k(c+r+2)-T_k(c+r+1)\bigr]\\
        &\quad=\Delta_k^{(1)}(c,r)
          +\Delta_k^{(1)}(c,r+1)\\
        &\quad=\Delta_k^{(2)}(c,r).
    \end{aligned}
    \label{eq:proof-pair-increment}
\end{equation}
The configurations \(01\) and \(10\) are exchanged without changing
the leaf Hamming weight, so each contributes unity.  The sum over the four bit
configurations is therefore
\begin{equation}
    2+2\cos[\phi\Delta_k^{(2)}(c,r)].
\end{equation}
Combining this with the multiplicity \(\binom{N-3}{r}\), summing over
\(c,r\), and multiplying by \(2^{-N}\) proves
Eq.~\eqref{eq:cutoff-C}.  This completes the derivation.

\paragraph*{Consistency limits.}
At \(\phi=0\), the binomial sums give
\((\mathsf{A}_{N,k},\mathsf{B}_{N,k},\mathsf{C}_{N,k})=(1,0,1)\);
thus \(\mathcal{F}_{Q}(0,k,N)=1\). Integer-valued phase counts and the
cosine representation also establish \(2\pi\)-periodicity and evenness in
\(\phi\).  

For \(N\geq3\), when \(k=2\), the sum defining \(D_k(s)\)
is empty, so \(D_2(s)=\sum_{q=2}^{1}\binom{s}{q}=0\) for
every \(s\). Consequently,
\(\Delta_2^{(1)}(c,r)=c\) and
\(\Delta_2^{(2)}(c,r)=2c\), independently of \(r\).
Using \(\sum_{r=0}^{n}\binom{n}{r}=2^n\), the sums over
\(r\) in Eqs.~\eqref{eq:cutoff-A} and \eqref{eq:cutoff-B}
each give \(2^{N-2}\); with the prefactor \(2^{1-N}\),
this leaves \(1/2\). Their sums over \(c=0,1\) give
\(1+\cos\phi\) and \(1-\cos\phi\), respectively.
For Eq.~\eqref{eq:cutoff-C}, the sum over \(r\) gives
\(2^{N-3}\), leaving a prefactor of \(1/4\). Its sum over
\(c\) gives
\((1+\cos 0)+(1+\cos 2\phi)=3+\cos(2\phi)\).
Therefore,
\begin{equation}
    \begin{aligned}
        \mathsf{A}_{N,2}&=\frac{1+\cos\phi}{2},
        &\mathsf{B}_{N,2}&=\frac{1-\cos\phi}{2},\\
        \mathsf{C}_{N,2}&=\frac{3+\cos(2\phi)}{4}.
    \end{aligned}
    \label{eq:proof-star-correlators}
\end{equation}
Eq.~\eqref{eq:cutoff-qfi} then becomes
\begin{equation}
    \begin{aligned}
        \mathcal{F}_{Q}(\phi,2,N)
        ={}&1+4(N-1)\sin^2\!\left(\frac{\phi}{2}\right)\\
        &+(N-1)(N-3)\sin^4\!\left(\frac{\phi}{2}\right),
    \end{aligned}
    \label{eq:proof-star-sine}
\end{equation}
which is equivalent to Eq.~\eqref{eq:star-qfi} and yields
\(\mathcal{F}_{Q}=N^2\) at \(\phi=\pi\).  Since the spectral range of
\(\hat{H}_{\mathrm{enc}}\) is \(N\), this saturates the pure-state bound
\(\mathcal{F}_{Q}\leq
(\lambda_{\max}-\lambda_{\min})^2=N^2\) (see Ref.~\cite{pezze2018quantum}).

% For \(k=N\), the arguments required by the one- and two-leaf
% increments are \(s=c+r\) and \(s=c+r+1\), respectively. Their
% summation ranges give \(0\leq s\leq N-1\). Since
% \(\binom{s}{q}=0\) for \(q>s\), the binomial theorem gives
% \[
% \begin{aligned}
% D_N(s)
% &=\sum_{q=2}^{N-1}\binom{s}{q}\\
% &=\sum_{q=0}^{s}\binom{s}{q}
%   -\binom{s}{0}-\binom{s}{1}\\
% &=2^s-1-s.
% \end{aligned}
% \]
% Substituting this result into
% \(\Delta_N^{(1)}(c,r)=c+D_N(c+r)\), and adding successive
% one-leaf increments for \(\Delta_N^{(2)}\), yields
% \begin{equation}
% \begin{aligned}
% \Delta_N^{(1)}(c,r)
% &=c+\bigl[2^{c+r}-1-(c+r)\bigr]\\
% &=2^{c+r}-(r+1),\\
% \Delta_N^{(2)}(c,r)
% &=\Delta_N^{(1)}(c,r)
%   +\Delta_N^{(1)}(c,r+1)\\
% &=\bigl[2^{c+r}-(r+1)\bigr]\\
% &\quad+\bigl[2^{c+r+1}-(r+2)\bigr]\\
% &=3\cdot2^{c+r}-(2r+3).
% \end{aligned}
% \label{eq:proof-complete-increments}
% \end{equation}

\subsection{Adaptation to the leaf-only class}
\label{app:leaf-only-qfi-proof}
The four-qubit class \(\bm{\sigma}_4=(0,1,0)\) is not in the cutoff family
\(E_{N,k}\). Nevertheless, the same proof method applies because this
probe has equal-magnitude amplitudes and leaf-permutation symmetry.
Its phase count from Eq.~\eqref{eq:n4-phase-count} is
\begin{equation}
g_{\mathrm{leaf}}^{(4)}(c,w)
=g_{(0,1,0)}(c,w)
=cw+\binom{w}{3}.
\label{eq:proof-leaf-only-phase}
\end{equation}
The \(N\)-qubit generalization evaluated in
Appendix~\ref{app:critical-scaling} retains exactly one full-leaf hyperedge
\(L=\{2,\ldots,N\}\), giving
\begin{equation}
g_{\mathrm{leaf}}^{(N)}(c,w)
=cw+\binom{w}{N-1}
=cw+\delta_{w,N-1},
\label{eq:proof-full-leaf-phase}
\end{equation}
where \(0\leq w\leq N-1\) and \(\delta_{w,N-1}\) is the Kronecker
delta. Thus no lower-order leaf-only orbits
\(B_3,\ldots,B_{N-2}\) are included when \(N>4\).

Choose one leaf \(j\) to flip, and let \(r\) count the occupied leaves
among the other \(N-2\) leaves. The two strings paired by \(\hat X_j\)
have leaf Hamming weights \(r\) and \(r+1\), so their phase-count difference is
\begin{equation}
\begin{aligned}
&g_{\mathrm{leaf}}^{(N)}(c,r+1)
-g_{\mathrm{leaf}}^{(N)}(c,r)\\
&\quad=c+\delta_{r+1,N-1}-\delta_{r,N-1}\\
&\quad=c+\delta_{r,N-2},
\qquad 0\leq r\leq N-2.
\end{aligned}
\label{eq:proof-full-leaf-single-increment}
\end{equation}
For \(r<N-2\), the full-leaf hyperedge remains inactive after the flip,
and the increment is just \(c\). At \(r=N-2\), every other leaf is
occupied, so flipping \(j\) also activates the full-leaf hyperedge and
adds one to the increment. There are \(\binom{N-2}{r}\) assignments of
the other leaves with Hamming weight \(r\). Consequently, the ordinary
assignments have total multiplicity
\begin{equation}
\sum_{r=0}^{N-3}\binom{N-2}{r}=2^{N-2}-1,
\label{eq:proof-full-leaf-ordinary-multiplicity}
\end{equation}
whereas the exceptional assignment at \(r=N-2\) occurs once.

\begin{figure*}[!t]
  \centering
  \panelgraphic[width=0.7\textwidth]{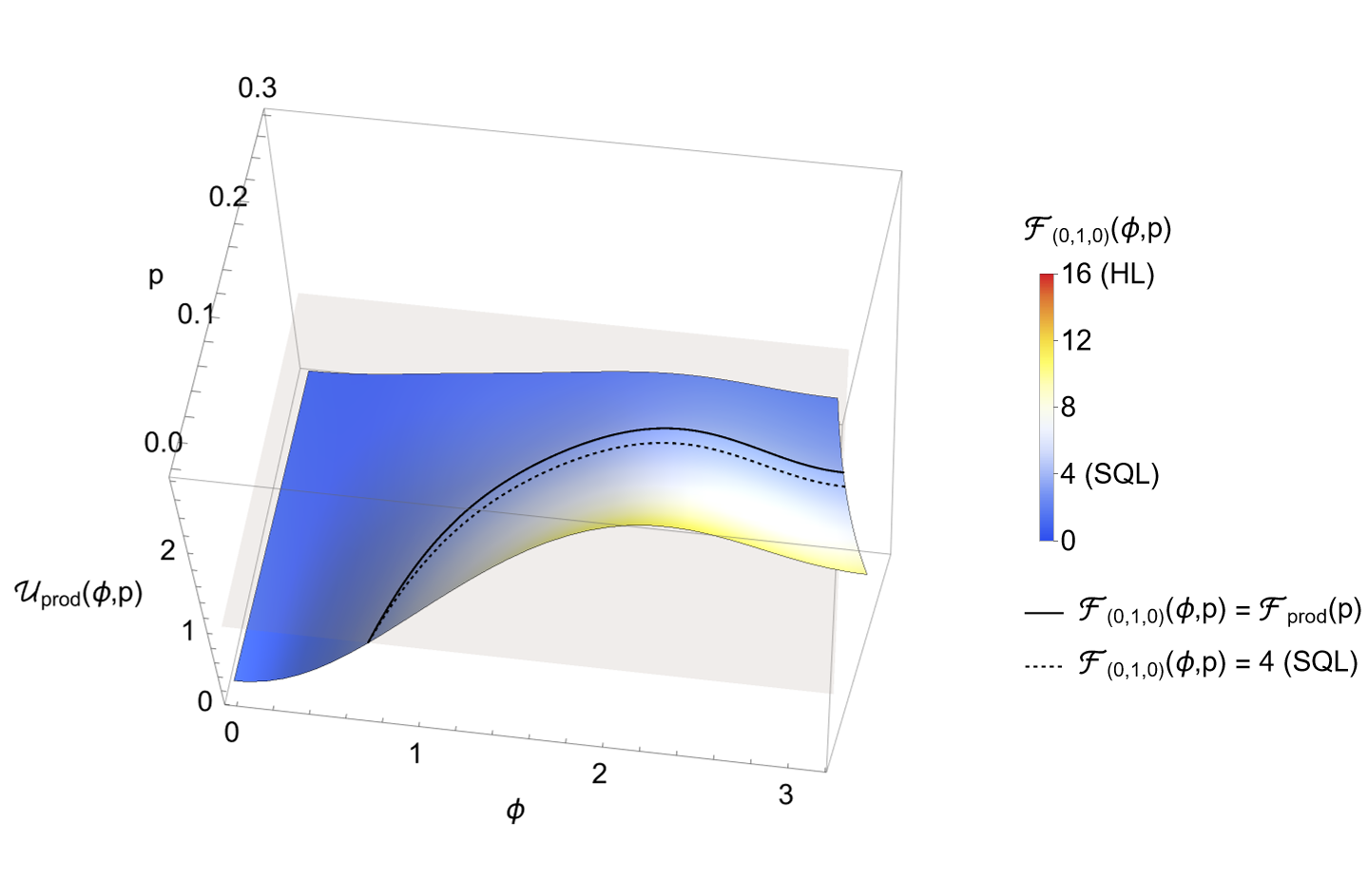}{\panellabel{8}{66.74}{a}} \\[0.em]
  \panelgraphic[width=0.7\textwidth]{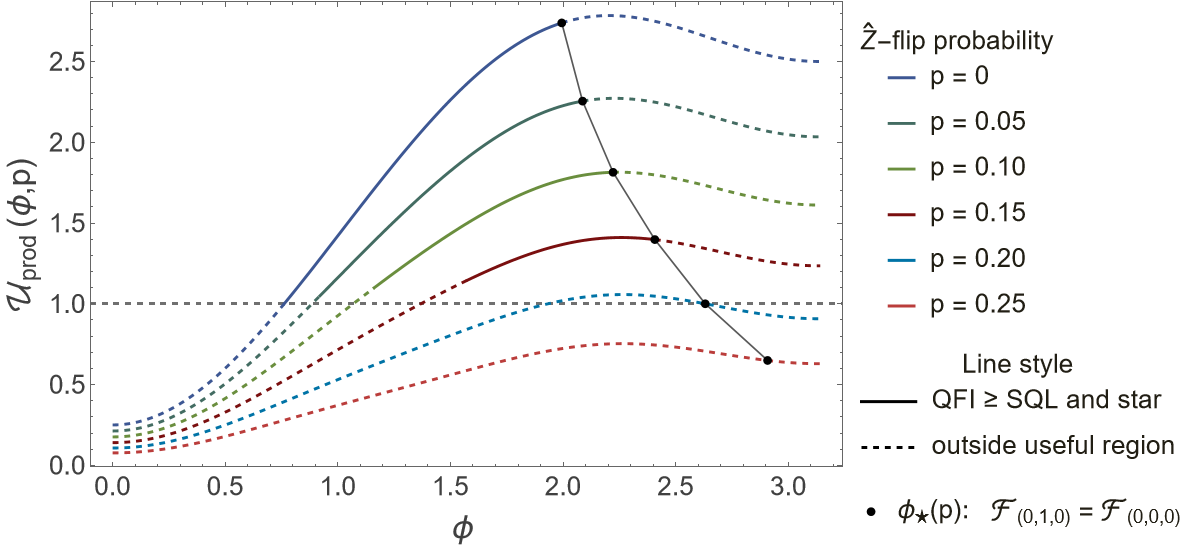}{\panellabel{8}{44.5}{b}}
  \caption{The product-state utility for
  $\bm{\sigma}_4 = (0,1,0)$: single full-leaf hyperedge.
  (a) Utility surface, with absolute QFI encoded by color.
  (b) Fixed-$p$ sections versus the interaction weight.
  Unity denotes equality with the dephased product-state QFI.
  In panel~(b), solid portions meet or exceed
  both the SQL and the equally dephased weighted star QFI, while dashed
  portions fail at least one comparison. Black points mark QFI crossings with the weighted star at nonzero weights, including below the SQL.}
  \label{fig:product-utility-010}
\end{figure*}

\begin{figure*}[!t]
  \centering
  \panelgraphic[width=0.7\textwidth]{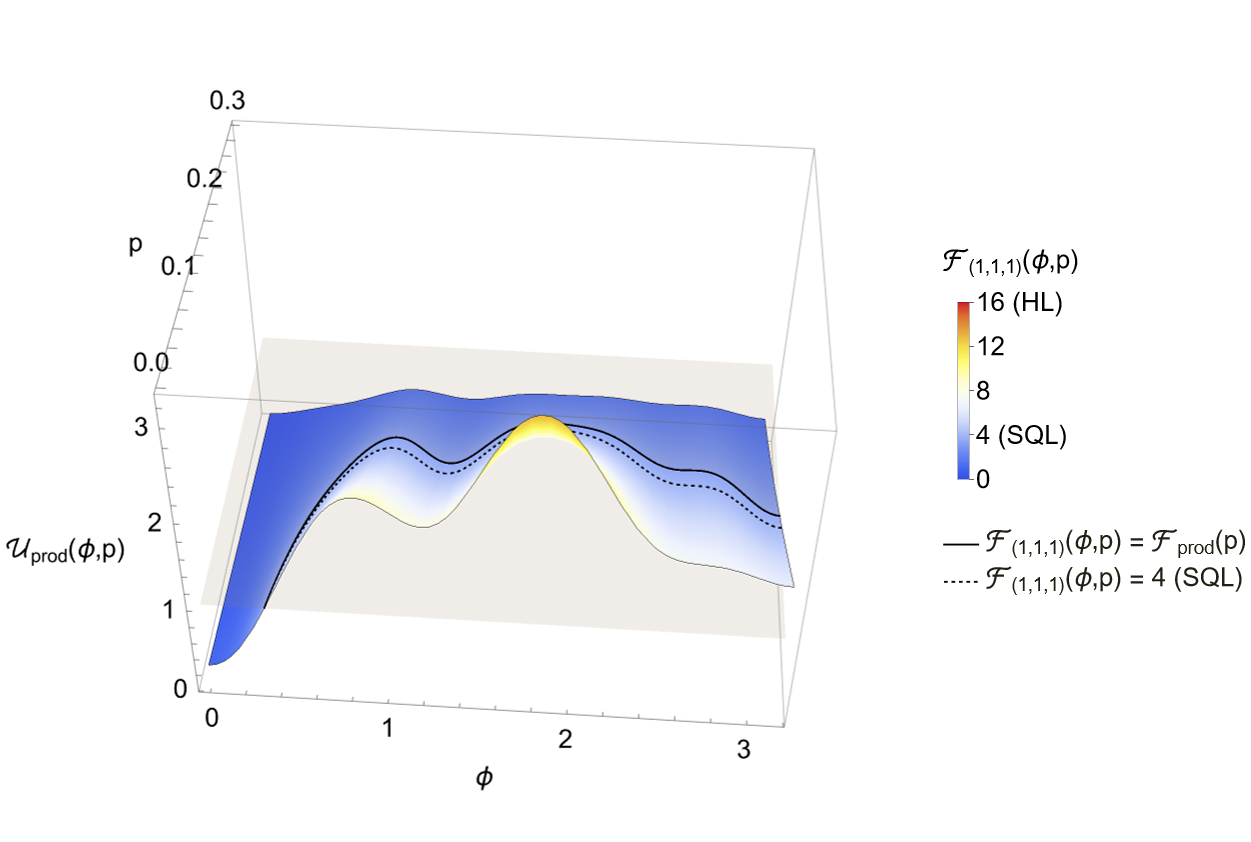}{\panellabel{8}{66.68}{a}} \\[0.0em]
\panelgraphic[width=0.7\textwidth]{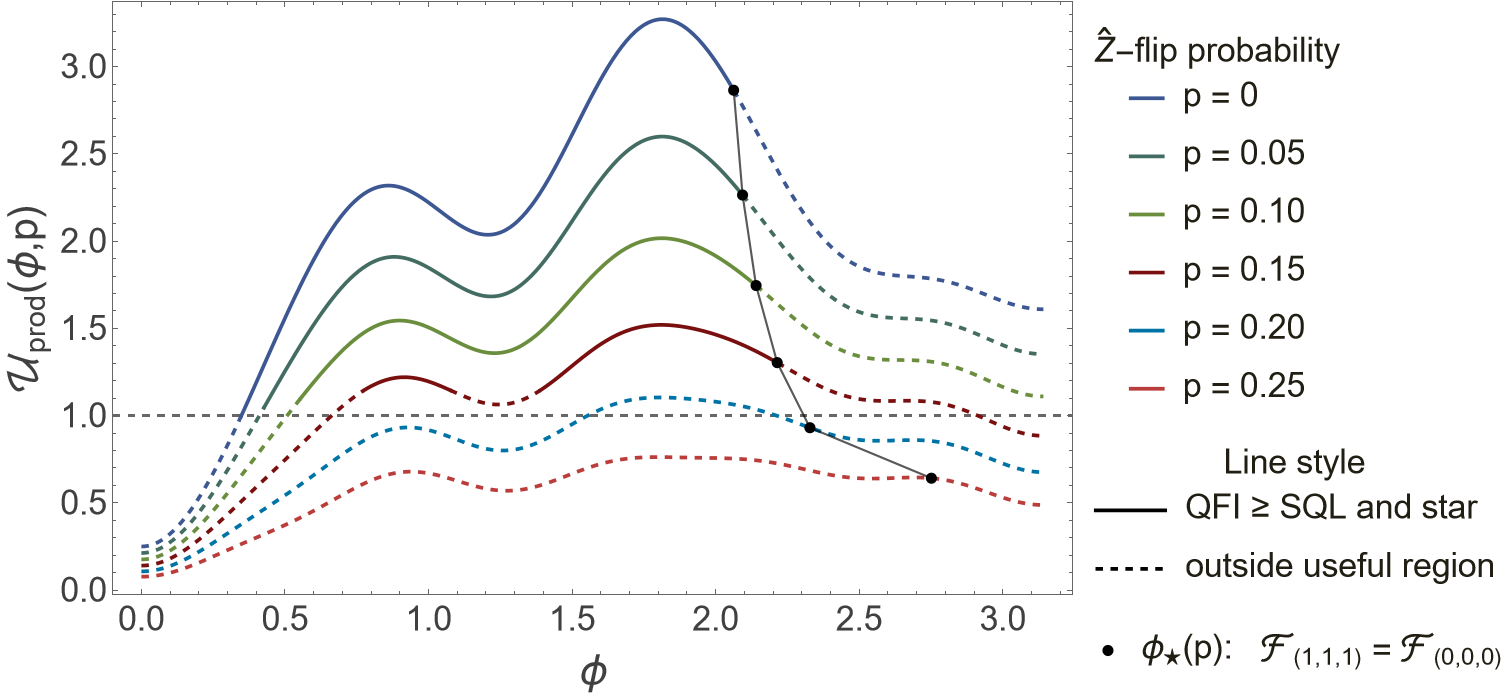}{\panellabel{8}{44.5}{b}}
  \caption{The product-state utility for
  $\bm{\sigma}_4 = (1,1,1)$: all higher-order hyperedges.
  (a) Utility surface, with absolute QFI encoded by color.
  (b) Fixed-$p$ sections versus the interaction weight.
  Unity denotes equality with the dephased product-state QFI.
  In panel~(b), solid portions meet or exceed
  both the SQL and the equally dephased weighted star QFI, while dashed
  portions fail at least one comparison. Black points mark QFI crossings with the weighted star at nonzero weights, including below the SQL.}
  \label{fig:product-utility-311}
\end{figure*}

For two distinct leaves \(j\) and \(\ell\), let \(r\) count the
occupied leaves among the remaining \(N-3\) leaves. There are
\(\binom{N-3}{r}\) assignments with this value of \(r\). The
\(00\leftrightarrow11\) pair has phase-count difference
\begin{equation}
g_{\mathrm{leaf}}^{(N)}(c,r+2)
-g_{\mathrm{leaf}}^{(N)}(c,r)
=2c+\delta_{r,N-3}.
\label{eq:proof-full-leaf-pair-increment}
\end{equation}
The \(01\leftrightarrow10\) pair preserves the leaf Hamming weight and has
zero phase-count difference. For fixed \(c\) and a fixed assignment
of the untouched leaves, these two pairs therefore contribute
\(2^{1-N}\cos[\phi(2c+\delta_{r,N-3})]\) and \(2^{1-N}\),
respectively, to \(\mathsf{C}_{N}^{\mathrm{leaf}}\). The extra
full-leaf phase in the first pair occurs only at \(r=N-3\), when
every untouched leaf is occupied. The ordinary assignments have
total multiplicity
\(\sum_{r=0}^{N-4}\binom{N-3}{r}=2^{N-3}-1\), and the exceptional
assignment occurs once.

Write \(u=\cos\phi\), \(v=\cos(2\phi)\), and
\(t=\cos(3\phi)\). For the one-leaf flip, let \(S_c\) be the
cosine sum for a fixed central bit \(c\), with the common factor
\(2^{1-N}\) omitted. The \(2^{N-2}-1\) ordinary assignments have
\(r<N-2\), while the exceptional assignment \(r=N-2\) occurs
once. Hence
\begin{equation}
\begin{aligned}
S_0
&=\sum_{r=0}^{N-3}\binom{N-2}{r}\cos 0+\cos\phi\\
&=(2^{N-2}-1)+u,\\
S_1
&=\sum_{r=0}^{N-3}\binom{N-2}{r}\cos\phi+\cos(2\phi)\\
&=(2^{N-2}-1)u+v.
\end{aligned}
\label{eq:proof-full-leaf-cosine-sums}
\end{equation}
Thus \(\mathsf{A}_{N}^{\mathrm{leaf}}=2^{1-N}(S_0+S_1)\).
For \(\mathsf{B}_{N}^{\mathrm{leaf}}\), the \(c=1\) sum is
subtracted because \(\hat Z_1\) contributes \((-1)^c\), giving
\(\mathsf{B}_{N}^{\mathrm{leaf}}=2^{1-N}(S_0-S_1)\).

For the two-leaf flip, after taking out the common factor
\(2^{1-N}\), the \(01\leftrightarrow10\) pair contributes \(1\)
and the \(00\leftrightarrow11\) pair contributes
\(\cos[\phi(2c+\delta_{r,N-3})]\). For each ordinary assignment
\(r<N-3\), their sum is \(1+\cos 0=2\) at \(c=0\) and
\(1+\cos(2\phi)=1+v\) at \(c=1\). Summing over \(c=0,1\) gives \(3+v\) for each assignment of the remaining \(N-3\) leaves with \(r<N-3\). For the
exceptional assignment \(r=N-3\), the corresponding sums are
\(1+\cos\phi=1+u\) at \(c=0\) and
\(1+\cos(3\phi)=1+t\) at \(c=1\), giving \(2+u+t\).
Thus, for \(N\geq4\),
\begin{align}
\mathsf{A}_{N}^{\mathrm{leaf}}
&=\frac{(2^{N-2}-1)+2^{N-2}u+v}{2^{N-1}},\nonumber\\
\mathsf{B}_{N}^{\mathrm{leaf}}
&=\frac{(2^{N-2}-1)(1-u)+u-v}{2^{N-1}},\nonumber\\
\mathsf{C}_{N}^{\mathrm{leaf}}
&=\frac{(2^{N-3}-1)(3+v)+2+u+t}{2^{N-1}}.
\label{eq:proof-leaf-only-correlators}
\end{align}
Here \(\mathsf{A}=\langle\hat X_j\rangle\),
\(\mathsf{B}=\langle\hat Z_1\hat X_j\rangle\), and
\(\mathsf{C}=\langle\hat X_j\hat X_\ell\rangle\) for distinct leaves
\(j,\ell\). The corresponding noiseless QFI under \(\hat H_\star\) is
\begin{equation}
\begin{aligned}
\mathcal{F}_{Q}^{\mathrm{leaf}}(N,\phi)
={}&N+2(N-1)\mathsf{B}_{N}^{\mathrm{leaf}}\\
&+(N-1)(N-2)\mathsf{C}_{N}^{\mathrm{leaf}}\\
&-(N-1)^2(\mathsf{A}_{N}^{\mathrm{leaf}})^2.
\end{aligned}
\label{eq:proof-leaf-only-qfi}
\end{equation}
This family is distinct from the cutoff family \(E_{N,k}\).

\section{Optimal separable benchmarks and product-state utility plots}
\label{app:additional-utility}

\paragraph*{Optimality of the dephased separable benchmark.}
We show that Eq.~\eqref{eq:product-qfi} is the maximum QFI over fully
separable inputs when the channel \(\mathcal E_p^{\otimes N}\) in
Eq.~\eqref{depchan} acts before encoding with
\(\hat U_\theta=e^{-i\theta\hat H_\star}\), as in
Eq.~\eqref{eq:noisy-encoded-state}. Here \(\hat H_\star\) is defined in
Eq.~\eqref{eq:encoding-generator}. An arbitrary fully separable input is
\begin{equation}
    \rho_{\mathrm{sep}}
    =\sum_\lambda q_\lambda\bigotimes_{j=1}^N\rho_j^{(\lambda)},
    \qquad q_\lambda\geq0,
    \quad\sum_\lambda q_\lambda=1.
    \label{eq:separable-decomposition}
\end{equation}
Let \(\rho_{\mathrm{sep},\theta}\) denote the encoded output. The channel and
encoding act locally, so each component remains a product. Convexity of the QFI and
additivity for product states under
a sum of local generators~\cite{toth2014quantum,liu2020quantum} yield
\begin{equation}
    \mathcal F_Q(\rho_{\mathrm{sep},\theta})
    \leq\sum_\lambda q_\lambda\sum_{j=1}^N
    \mathcal F_Q\!\left(\rho_{j,\theta}^{(\lambda)}\right),
    \label{eq:separable-qfi-convexity}
\end{equation}

where
\(\rho_{j,\theta}^{(\lambda)}=
e^{-i\theta\hat h_j}\mathcal E_p(\rho_j^{(\lambda)})e^{i\theta\hat h_j}\),
\(\hat h_1=\hat Z_1/2\), and \(\hat h_j=\hat X_j/2\) for \(j\geq2\).

For a single-qubit input state \(\rho\) before dephasing, define its
Bloch vector by
\(\bm s=(s_x,s_y,s_z)
=(\langle\hat X\rangle_\rho,\langle\hat Y\rangle_\rho,
\langle\hat Z\rangle_\rho)\).
Then
\(\rho=\tfrac12(\hat{\mathbb I}+s_x\hat X+s_y\hat Y+s_z\hat Z)\),
with \(|\bm s|\leq1\) (see Ref.~\cite{liu2020quantum}).
The phase-flip channel in Eq.~\eqref{eq:single-dephasing-channel} gives
\begin{equation}
    \bm r=(\eta s_x,\eta s_y,s_z),
    \qquad \eta=1-2p.
    \label{eq:dephased-product-bloch-vector}
\end{equation}
For \(\hat h=\bm n\cdot\hat{\bm\sigma}/2\), with \(|\bm n|=1\) and
\(\hat{\bm\sigma}=(\hat X,\hat Y,\hat Z)\), the qubit Bloch-vector
formula gives \(\mathcal F_Q=|\bm n\times\bm r|^2\) (see Ref.~\cite{liu2020quantum}):
unitary encoding has \(\partial_\theta\bm r_\theta=\bm n\times\bm r_\theta\),
so its radial QFI term vanishes. The formula also holds for pure states.
The central and leaf contributions are therefore
\begin{align}
    \mathcal F_Q^{(1)}
    &=\eta^2(s_{1x}^2+s_{1y}^2)\leq\eta^2,
    \label{eq:optimal-central-qfi}\\
    \mathcal F_Q^{(j)}
    &=\eta^2s_{jy}^2+s_{jz}^2\leq1,
    \qquad j=2,\ldots,N.
    \label{eq:optimal-leaf-qfi}
\end{align}

Thus Eq.~\eqref{eq:separable-qfi-convexity} bounds every fully separable
input by \(N-1+\eta^2\). The product state in
Eq.~\eqref{eq:product-reference} attains both local bounds: its central
input Bloch vector is \((1,0,0)\) and each leaf has \((0,0,1)\). Hence
\begin{equation}
    \max_{\rho_{\mathrm{sep}}}
    \mathcal F_Q\!\left[
        \hat U_\theta\mathcal E_p^{\otimes N}(\rho_{\mathrm{sep}})
        \hat U_\theta^\dagger
    \right]
    =N-1+(1-2p)^2.
    \label{eq:optimal-separable-qfi}
\end{equation}
The product input in Eq.~\eqref{eq:product-reference} saturates this
bound for every \(0\leq p\leq1/2\), although it is not unique.

For the complete-hypergraph generator in
Eq.~\eqref{eq:complete-encoding-generator}, each single-qubit QFI is at most
one. The input \(\ket{0}^{\otimes N}\) is unchanged by the channel and
attains this bound on every qubit. Thus
\begin{equation}
    \mathcal F_{\mathrm{sep,opt}}^{\mathrm{comp}}(N,p)=N,
    \qquad 0\leq p\leq\tfrac12.
    \label{eq:complete-separable-bound}
\end{equation}

Figs.~\ref{fig:product-utility-010} and~\ref{fig:product-utility-311}
supplement the star-relative utility in Eq.~\eqref{eq:star-utility}
with the product-state utility of Eq.~\eqref{eq:product-utility} for
\(\bm\sigma_4=(0,1,0)\) and \((1,1,1)\), respectively. Both use
\(\hat H_\star\), so a
product-state ratio above unity is an advantage over the optimal
dephased separable benchmark. In each figure, panel~(a) shows the utility surface;
panel~(b) shows sections at \(p=0,0.05,0.10,0.15,0.20,0.25\).

For the fixed-\(p\) sections in panel (b), \(\phi_{\mathrm{SQL}}(p)\) is the first upward SQL
crossing when one exists, and \(\phi_\star(p)\) denotes a nonzero crossing
with the equally dephased weighted star graph state; several such crossings may occur at a given \(p\). Solid segments meet both the SQL, \(\mathcal F_Q=4\), and the equally dephased weighted star QFI; dashed segments fail at least one condition. The qualifying weights may be disconnected, as for
\(\bm\sigma_4=(1,1,1)\) at \(p=0.15\). Black points mark
\(\phi_\star(p)\) even where the SQL is not reached.

\section{Numerically optimized critical-probability scaling}
\label{app:critical-scaling}

For $N>4$, we generalize the three four-qubit star-based families using the edge sets
\begin{align}
    E_{\mathrm{star}}^{(N)}
    &=E_{\star},\\
    E_{\mathrm{all}}^{(N)}
    &=E_{N,N},\\
    E_{\mathrm{leaf}}^{(N)}
    &=E_{\star}\cup\{L\}.
\end{align}

Fig.~\ref{fig:critical-scaling} compares these families with the
complete weighted hypergraph of
Eq.~\eqref{eq:complete-hypergraph-edges} for even
\(N=4,6,\ldots,24\). The star-based families use
\(\hat H_{\star}\) from Eq.~\eqref{eq:encoding-generator}, whereas
the complete family uses \(\hat H_{\mathrm{comp}}\) from
Eq.~\eqref{eq:complete-encoding-generator}. We evaluate the dephased state's QFI using
Eq.~\eqref{eq:mixed-qfi-spectral}. For each probe family and fixed
\(\phi\), we determine the critical dephasing probability at which
the QFI first falls to the chosen benchmark as \(p\) increases from
zero, as defined in Eq.~\eqref{eq:pointwise-critical-probabilities}.
We then maximize this critical probability over \(\phi\in[0,\pi]\),
as in Eq.~\eqref{eq:optimized-critical-probabilities}.
This optimization is performed separately for the SQL and the
shared product-state reference
\(\mathcal F_{\mathrm{prod}}(N,p)=N-1+(1-2p)^2\), so the two
benchmarks may yield different optimal weights.

For each fixed \(N\) and \(\phi\), we compared the QFI with the
chosen benchmark at 121 equally spaced values of \(p\in[0,1/2]\).
We then used bisection to narrow the first crossing interval to a
width below \(10^{-12}\). We refined the \(\phi\) grids to resolve narrow maxima in the critical-probability curves that could be missed by coarser sampling. Starting with 80 equal subintervals of \([0,\pi]\), we doubled the number of subintervals at each \(\phi\)-grid refinement. The finest grids used 655,360 subintervals for the weighted star and single-full-leaf families, and 41,943,040 for the all-higher-order and complete families at \(N=24\). Grid searches were followed
by local refinement of candidate maxima. All 88 threshold estimates
changed by less than \(10^{-7}\) over the final two \(\phi\)-grid
refinements. These checks support numerical stability of the
estimates but do not certify a global maximum over
continuous \(\phi\).

\FloatBarrier
\flushbottom


\begin{thebibliography}{99}
\interlinepenalty=10000

\bibitem{horodecki2009quantum}
R.~Horodecki, P.~Horodecki, M.~Horodecki and K.~Horodecki,
``Quantum entanglement'',
\textit{Reviews of Modern Physics}, \textbf{81}, 865, (2009).

\bibitem{giovannetti2011advances}
V.~Giovannetti, S.~Lloyd and L.~Maccone,
``Advances in quantum metrology'',
\textit{Nature Photonics}, \textbf{5}, 222, (2011).

\bibitem{toth2014quantum}
G. T\'oth and I. Apellaniz,
``Quantum metrology from a quantum information science perspective'',
\textit{Journal of Physics A: Mathematical and Theoretical}, \textbf{47}, 424006, (2014).

\bibitem{liu2020quantum}
J.~Liu, H.~Yuan, X.-M.~Lu and X.~Wang,
``Quantum Fisher information matrix and multiparameter estimation'',
\textit{Journal of Physics A: Mathematical and Theoretical}, \textbf{53}, 023001, (2020).

\bibitem{degen2017quantum}
C. L. Degen, F. Reinhard and P. Cappellaro,
``Quantum sensing'',
\textit{Reviews of Modern Physics}, \textbf{89}, 035002, (2017).

\bibitem{pezze2018quantum}
L. Pezz\`e, A. Smerzi, M. K. Oberthaler, R. Schmied and
P. Treutlein,
``Quantum metrology with nonclassical states of atomic ensembles'',
\textit{Reviews of Modern Physics}, \textbf{90}, 035005, (2018).

\bibitem{montenegro2025review}
V. Montenegro, C. Mukhopadhyay, R. Yousefjani, S. Sarkar,
U. Mishra, M. G. A. Paris and A. Bayat,
``Review: Quantum metrology and sensing with many-body systems'',
\textit{Physics Reports}, \textbf{1134}, 1, (2025).

\bibitem{greenberger1989going}
D.~M. Greenberger, M.~A. Horne and A. Zeilinger,
``Going Beyond Bell's Theorem'',
in \textit{Bell's Theorem, Quantum Theory and Conceptions of the Universe},
edited by M. Kafatos
(Springer, Dordrecht, 1989), pp.~69--72.

\bibitem{bollinger1996optimal}
J.~J.~Bollinger, W.~M.~Itano, D.~J.~Wineland and D.~J.~Heinzen,
``Optimal frequency measurements with maximally correlated states'',
\textit{Physical Review A}, \textbf{54}, R4649, (1996).

\bibitem{boto2000quantum}
A.~N. Boto, P. Kok, D.~S. Abrams, S.~L. Braunstein,
C.~P. Williams and J.~P. Dowling,
``Quantum interferometric optical lithography: Exploiting entanglement to beat the diffraction limit'',
\textit{Physical Review Letters}, \textbf{85}, 2733, (2000).

\bibitem{wineland1992spin}
D. J. Wineland, J. J. Bollinger, W. M. Itano, F. L. Moore
and D. J. Heinzen,
``Spin squeezing and reduced quantum noise in spectroscopy'',
\textit{Physical Review A}, \textbf{46}, R6797, (1992).

\bibitem{kitagawa1993squeezed}
M. Kitagawa and M. Ueda,
``Squeezed spin states'',
\textit{Physical Review A}, \textbf{47}, 5138, (1993).

% \bibitem{caves1981quantum}
% C.~M. Caves,
% ``Quantum-mechanical noise in an interferometer'',
% \textit{Physical Review D}, \textbf{23}, 1693, (1981).

% \bibitem{holland1993interferometric}
% M.~J. Holland and K. Burnett,
% ``Interferometric detection of optical phase shifts at the Heisenberg limit'',
% \textit{Physical Review Letters}, \textbf{71}, 1355, (1993).

\bibitem{huelga1997improvement}
S. F. Huelga, C. Macchiavello, T. Pellizzari, A. K. Ekert,
M. B. Plenio and J. I. Cirac,
``Improvement of frequency standards with quantum entanglement'',
\textit{Physical Review Letters}, \textbf{79}, 3865, (1997).

\bibitem{demkowicz2012elusive}
R. Demkowicz-Dobrza\'nski, J. Ko{\l}ody\'nski and M. Gu\c{t}\u{a},
``The elusive Heisenberg limit in quantum-enhanced metrology'',
\textit{Nature Communications}, \textbf{3}, 1063, (2012).

\bibitem{alexander2020generating}
B. Alexander, J.~J. Bollinger and H. Uys,
``Generating Greenberger--Horne--Zeilinger states with squeezing and postselection'',
\textit{Physical Review A}, \textbf{101}, 062303, (2020).

\bibitem{alexander2024robustness}
B.~J. Alexander, J.~J. Bollinger and M.~S. Tame,
``Robustness of the projected squeezed state protocol'',
\textit{Physical Review A}, \textbf{109}, 052614, (2024).

\bibitem{linnemann2016quantum}
D.~Linnemann, H.~Strobel, W.~Muessel, J.~Schulz,
R.~J. Lewis-Swan, K.~V. Kheruntsyan and M.~K. Oberthaler,
``Quantum-enhanced sensing based on time reversal of nonlinear dynamics'',
\textit{Physical Review Letters}, \textbf{117}, 013001, (2016).

\bibitem{dorner2012quantum}
U.~Dorner,
``Quantum frequency estimation with trapped ions and atoms'',
\textit{New Journal of Physics}, \textbf{14}, 043011, (2012).

\bibitem{roos2006designer}
C. F. Roos, M. Chwalla, K. Kim, M. Riebe and R. Blatt,
``\,`Designer atoms' for quantum metrology'',
\textit{Nature}, \textbf{443}, 316, (2006).

\bibitem{kessler2014quantum}
E. M. Kessler, I. Lovchinsky, A. O. Sushkov and M. D. Lukin,
``Quantum error correction for metrology'',
\textit{Physical Review Letters}, \textbf{112}, 150802, (2014).

\bibitem{arrad2014increasing}
G. Arrad, Y. Vinkler, D. Aharonov and A. Retzker,
``Increasing sensing resolution with error correction'',
\textit{Physical Review Letters}, \textbf{112}, 150801, (2014).

\bibitem{dur2014improved}
W. D{\"u}r, M. Skotiniotis, F. Fr{\"o}wis and B. Kraus,
``Improved quantum metrology using quantum error correction'',
\textit{Physical Review Letters}, \textbf{112}, 080801, (2014).

\bibitem{ozeri2013heisenberg}
R. Ozeri,
``Heisenberg limited metrology using quantum error-correction codes'',
arXiv:1310.3432 [quant-ph], (2013).

\bibitem{leibfried2004toward}
D. Leibfried, M. D. Barrett, T. Schaetz, J. Britton,
J. Chiaverini, W. M. Itano, J. D. Jost, C. Langer
and D. J. Wineland,
``Toward Heisenberg-limited spectroscopy with multiparticle
entangled states'',
\textit{Science}, \textbf{304}, 1476, (2004).

\bibitem{gilmore2021quantum}
K. A. Gilmore, M. Affolter, R. J. Lewis-Swan, D. Barberena,
E. Jordan, A. M. Rey and J. J. Bollinger,
``Quantum-enhanced sensing of displacements and electric fields
with two-dimensional trapped-ion crystals'',
\textit{Science}, \textbf{373}, 673, (2021).

\bibitem{marciniak2022optimal}
C.~D. Marciniak, T. Feldker, I. Pogorelov, R. Kaubruegger,
D.~V. Vasilyev, R. van Bijnen, P. Schindler, P. Zoller,
R. Blatt and T. Monz,
``Optimal metrology with programmable quantum sensors'',
\textit{Nature}, \textbf{603}, 604, (2022).

\bibitem{gross2010nonlinear}
C. Gross, T. Zibold, E. Nicklas, J. Est\`eve and
M. K. Oberthaler,
``Nonlinear atom interferometer surpasses classical precision limit'',
\textit{Nature}, \textbf{464}, 1165, (2010).

\bibitem{finkelstein2024universal}
R. Finkelstein, R. B.-S. Tsai, X. Sun, P. Scholl, S. Direkci,
T. Gefen, J. Choi, A. L. Shaw and M. Endres,
``Universal quantum operations and ancilla-based read-out for
tweezer clocks'',
\textit{Nature}, \textbf{634}, 321, (2024).

\bibitem{nagata2007beating}
T. Nagata, R. Okamoto, J.~L. O'Brien, K. Sasaki and S. Takeuchi,
``Beating the standard quantum limit with four-entangled photons'',
\textit{Science}, \textbf{316}, 726, (2007).

\bibitem{slussarenko2017unconditional}
S. Slussarenko, M. M. Weston, H. M. Chrzanowski, L. K. Shalm,
V. B. Verma, S. W. Nam and G. J. Pryde,
``Unconditional violation of the shot-noise limit in photonic
quantum metrology'',
\textit{Nature Photonics}, \textbf{11}, 700, (2017).

\bibitem{xu2022metrological}
K. Xu et al.,
``Metrological characterization of non-Gaussian entangled states
of superconducting qubits'',
\textit{Physical Review Letters}, \textbf{128}, 150501, (2022).

\bibitem{zhou2020quantum}
H. Zhou et al.,
``Quantum metrology with strongly interacting spin systems'',
\textit{Physical Review X}, \textbf{10}, 031003, (2020).

\bibitem{gao2025signal}
H. Gao et al.,
``Signal amplification in a solid-state sensor through asymmetric
many-body echo'',
\textit{Nature}, \textbf{646}, 68, (2025).

\bibitem{gachechiladze2016extreme}
M. Gachechiladze, C. Budroni and O. G\"uhne,
``Extreme violation of local realism in quantum hypergraph states'',
\textit{Physical Review Letters}, \textbf{116}, 070401, (2016).

\bibitem{alexander2026weighted}
B. J. Alexander, \c{S}. K. \"{O}zdemir and M. S. Tame,
``Weighted graph states as a resource for quantum metrology'',
arXiv:2602.13026 [quant-ph], (2026).

\bibitem{skosana2026experimental}
U. Skosana, B. Alexander, C. Lee and M. Tame,
``Experimental realization of a photonic weighted graph state
for quantum metrology'',
arXiv:2602.18177 [quant-ph], (2026).

\bibitem{shettell2020graph}
N. Shettell and D. Markham,
``Graph states as a resource for quantum metrology'',
\textit{Physical Review Letters}, \textbf{124}, 110502, (2020).

\bibitem{gachechiladze2019thesis}
M. Gachechiladze,
\textit{Quantum Hypergraph States and the Theory of Multiparticle Entanglement},
Ph.D. thesis, University of Siegen (2019).
\url{https://doi.org/10.25819/ubsi/493}.

\bibitem{salem2025review}
V. Salem,
``Quantum hypergraph states: a review'',
arXiv:2512.02955 [quant-ph], (2025).

\bibitem{fajardo2026restriction}
C. Fajardo and M. Paraschiv,
``Restriction-based certificate of bipartite Schmidt rank in hypergraph states'',
\textit{Physical Review A}, \textbf{114}, 022410, (2026).

\bibitem{scarafile2026statistical}
P. Scarafile, G. Trotta, P. Facchi, G. Magnifico, G. Parisi,
S. Pascazio and K. \.{Z}yczkowski,
``Statistical mechanics of multipartite entanglement in hypergraph states'',
arXiv:2609.24508 [quant-ph], (2026).

\bibitem{susulovska2026three}
N. A. Susulovska and Kh. P. Gnatenko,
``Three-uniform edge-ordered hypergraph quantum states:
entanglement and its relation to hypergraph properties'',
arXiv:2609.30399 [quant-ph], (2026).

\bibitem{salem2026randomized}
V. Salem, A. A. Silva and F. M. Andrade,
``Randomized hypergraph states and their entanglement properties'',
\textit{Annalen der Physik}, \textbf{538}, e00622, (2026).

\bibitem{park2026lowoverhead}
G. Park, J. Sun and H. Jeong,
``Low-overhead tailoring and learning of noise in graph states'',
\textit{Physical Review A}, \textbf{114}, 042428, (2026).

\bibitem{jacobs2014quantum}
K.~Jacobs,
\textit{Quantum measurement theory and its applications},
(Cambridge University Press, 2014), p.~27.

\bibitem{hein2006entanglement}
M. Hein, W. D\"ur, J. Eisert, R. Raussendorf, M. Van den Nest
and H.-J. Briegel,
``Entanglement in graph states and its applications'',
in \textit{Quantum Computers, Algorithms and Chaos},
Proceedings of the International School of Physics
``Enrico Fermi,'' Vol.~162, edited by G. Casati,
D. L. Shepelyansky, P. Zoller and G. Benenti
(IOS Press, Amsterdam, 2006), pp.~115--218.


\bibitem{friis2017flexible}
N. Friis, D. Orsucci, M. Skotiniotis, P. Sekatski, V. Dunjko,
H. J. Briegel and W. D\"ur,
``Flexible resources for quantum metrology'',
\textit{New Journal of Physics}, \textbf{19}, 063044, (2017).

\bibitem{nguyen2024harnessing}
P. T. Nguyen, T. K. Le, H. Q. Nguyen and L. B. Ho,
``Harnessing graph state resources for robust quantum
magnetometry under noise'',
\textit{Scientific Reports}, \textbf{14}, 20528, (2024).

\bibitem{plodzien2025manybody}
M. P{\l}odzie\'n, M. Lewenstein and J. Chwede\'nczuk,
``Many-body quantum resources of graph states'',
\textit{Reports on Progress in Physics}, \textbf{88}, 077601, (2025).

\bibitem{cooper2024graph}
E. S. Cooper, P. Kunkel, A. Periwal and M. Schleier-Smith,
``Graph states of atomic ensembles engineered by
photon-mediated entanglement'',
\textit{Nature Physics}, \textbf{20}, 770, (2024).

\bibitem{qin2025scaling}
Z. Qin and V. W. Scarola,
``Scaling of computational order parameters in Rydberg-atom
graph states'',
\textit{Physical Review A}, \textbf{111}, 042617, (2025).

\bibitem{li2026deterministic}
Y.~Li, Y.~Yang, Z.~Bao, J.~Yang, L.~Sun, Y.~Song,
Y.~Wu, H.~Zhang and L.~Duan,
``Deterministic and programmable fusion for the scalable generation of photonic graph states'',
\textit{Nature Physics}, (2026).
\url{https://doi.org/10.1038/s41567-026-03471-5}.

\bibitem{hartmann2007weighted}
L.~Hartmann, J.~Calsamiglia, W.~D\"ur and H.~J.~Briegel,
``Weighted graph states and applications to spin chains, lattices and gases'',
\textit{Journal of Physics B: Atomic, Molecular and Optical Physics},
\textbf{40}, S1, (2007).

\bibitem{rimock2026type}
N.~Rimock and Y.~Oz,
``Type-I and type-II fusion protocols for weighted graph states'',
\textit{Physical Review A}, \textbf{114}, 042439, (2026).
\url{https://doi.org/10.1103/bf7h-2mk9}.

\bibitem{rossi2013quantum}
M. Rossi, M. Huber, D. Bru{\ss} and C. Macchiavello,
``Quantum hypergraph states'',
\textit{New Journal of Physics}, \textbf{15}, 113022, (2013).

\bibitem{guhne2014entanglement}
O. G\"uhne, M. Cuquet, F. E. S. Steinhoff, T. Moroder,
M. Rossi, D. Bru{\ss}, B. Kraus and C. Macchiavello,
``Entanglement and nonclassical properties of hypergraph states'',
\textit{Journal of Physics A: Mathematical and Theoretical}, \textbf{47}, 335303, (2014).

\bibitem{poderini2026hypergraph}
D. Poderini, D. Bru{\ss} and C. Macchiavello,
``Quantum hypergraph states: A review'',
\textit{Reports on Progress in Physics}, \textbf{89}, 066001, (2026).

\bibitem{arantes2026stabilizers}
G. M. Arantes, V. Salem, D. Cius and B. Amaral,
``k-uniform complete hypergraph-state stabilizers in terms of local operators'',
\textit{Physical Review A}, \textbf{113}, 052417, (2026).

\bibitem{zakaryan2025nonsymmetric}
H. Zakaryan, K.-R. Revis and Z. Raissi,
``Nonsymmetric Greenberger-Horne-Zeilinger states:
Weighted hypergraph and controlled-unitary graph representations'',
\textit{Physical Review A}, \textbf{112}, 032438, (2025).

\bibitem{huang2024demonstration}
J. Huang et al.,
``Demonstration of hypergraph-state quantum information processing'',
\textit{Nature Communications}, \textbf{15}, 2601, (2024).

\bibitem{bag2022achieving}
S. Bag, R. Sarkar and P. K. Panigrahi,
``Achieving Heisenberg limit in the phase measurement through
three-qubit graph states'',
arXiv:2208.07772 [quant-ph], (2022).

\bibitem{dicke1954coherence}
R. H. Dicke,
``Coherence in spontaneous radiation processes'',
\textit{Physical Review}, \textbf{93}, 99, (1954).

\bibitem{nielsen2002quantum}
M.~A. Nielsen and I.~L. Chuang,
\textit{Quantum Computation and Quantum Information: 10th Anniversary Edition}
(Cambridge University Press, Cambridge, 2010).

\bibitem{kim2025distributed}
D.-H.~Kim, S.~Hong, Y.-S.~Kim, K.~Oh, S.-Y.~Lee, C.~Lee and H.-T.~Lim,
``Distributed quantum sensing with multi-mode N00N states'',
\textit{Physical Review Letters}, \textbf{135}, 050802, (2025).

% \bibitem{nielsen2010quantum}
% M.~A.~Nielsen and I.~L.~Chuang,
% \textit{Quantum Computation and Quantum Information},
% 10th Anniversary ed. (Cambridge University Press, 2010).
\end{thebibliography}
\end{document}